\documentclass[final,3p,times,twocolumn]{elsarticle}

\usepackage{changes}
\usepackage{graphicx}
\usepackage{dcolumn}
\usepackage{bm}

\usepackage{amsmath,amssymb}
\usepackage[english]{babel}
\usepackage{graphicx}
\graphicspath{{./}{Figures/}}
\usepackage{xcolor}
\usepackage{cmap}
\usepackage{placeins}

\usepackage{hyperref}
\usepackage{soul}      
\usepackage{lipsum}    
\usepackage{blindtext} 
\usepackage{microtype} 

\definecolor{LightCyan}{rgb}{0.88,1,1}

\newcommand{\PWN}{\raise-0.4ex\hbox{\scalebox{0.8}{\scriptsize$P$\kern-0.05em$W$\kern-0.2em$N$}}}
\newcommand{\s}{\raise-0.1ex\hbox{\scalebox{1.2}{\scriptsize$s$}}}
\newcommand{\sla}{\;\raise0.55ex\hbox{\scriptsize$<$\kern-0.75em\raise-1.1ex\hbox{$\sim$}}\;}
\newcommand{\sga}{\;\raise0.55ex\hbox{\scriptsize$>$\kern-0.75em\raise-1.1ex\hbox{$\sim$}}\;}
\newcommand{\ssim}{\;\raise0.3ex\hbox{\tiny$\sim$}\,}
\newcommand{\sapprox}{\;\raise0.3ex\hbox{\tiny$\approx$}\,}

\newcommand{\porb}{\mbox {$P_{\mathrm orb}$}}
\def\lsim{\;\raise0.3ex\hbox{$<$\kern-0.75em\raise-1.1ex\hbox{$\sim$}}\;}
\def\gsim{\;\raise0.3ex\hbox{$>$\kern-0.75em\raise-1.1ex\hbox{$\sim$}}\;}

\def\kms{\rm ~km~s^{-1}}

\def \kms {\rm ~km~s$^{-1}$}

\def\ergs{\rm ~erg~s^{-1}}

\def\ecsb{erg cm$^{-2}$ s$^{-1}$ arcsec$^{-2}$ }
\def\ecsb2{erg cm$^{-2}$ s$^{-1}$ arcsec$^{-2}$}

\def\apj{ApJ}
\def\mnras{MNRAS}

\def\araa{ARA\&A}                
\def\aap{A\&A}                   
\def\aj{AJ}                      
\def\apjs{ApJS}                  
\def\apjl{ApJ}                   
\def\pasj{PASJ}

\def\ssr{Space Sci. Rev.}
\def\aapr{Astron. Astroph. Reviews}

\def\prl{Phys. Rev. Lett.}
\def\jcap{JCAP}
\def\prd{Phys. Rev. D}

\journal{Journal of High Energy Astrophysics}

\begin{document}

\begin{frontmatter}



\title{How many VHE gamma-ray binaries with young pulsars can be observed?}


\author[ioffe]{A. M. Bykov} 
\ead{byk@astro.ioffe.ru}
\author[gaish,ia]{A. G. Kuranov} 
\ead{alexandre.kuranov@gmail.com}
\author[ioffe]{A. E. Petrov} 
\ead{a.e.petrov@mail.ioffe.ru}
\author[gaish]{K. A. Postnov\corref{corKAP}} 
\ead{kpostnov@gmail.com}
\cortext[corKAP]{Corresponding author}

\affiliation[ioffe]{organization={Ioffe Institute},
            addressline={26, Politekhnicheskaya str.}, 
            city={St.~Petersburg},
            postcode={194021},
            country={Russia}}
            
\affiliation[gaish]{organization={Sternberg Astronomical Institute, M.V. Lomonosov Moscow State University},
            addressline={13 Universitetskij pr.}, 
            city={Moscow},
            postcode={119234}, 
            country={Russia}}

\affiliation[ia]{organization={Institute of Astronomy},
            addressline={48 Pyatnitskaya str.}, 
            city={Moscow},
            postcode={119017},
            country={Russia}}

\begin{abstract}
A population of Galactic gamma-ray  binaries is currently emerging due to ever increasing sensitivity of gamma-ray observatories. The detection of very high energy (VHE) photons with energies well above 10 TeV from a dozen of sources and the estimated power of those sources make them
potentially interesting cosmic ray accelerators.  Multi-wavelength observations of gamma-ray binaries revealed that most of them include a young massive star in pair with a relativistic companion, either a black hole or energetic pulsar. Fast stellar winds interacting with powerful relativistic outflows from  pulsars or the black hole jets in microquasars are favorable sites for very high energy particle acceleration. 
To estimate the expected number of gamma-ray binaries, we present here results of  population synthesis calculations  predicting the number of Galactic binaries in which a young massive OB- or Be-star is accompanied by a pulsar capable of producing a powerful relativistic outflow. The distributions over the binary eccentricities, orbital periods, Be-disk inclinations, and the pulsar braking energy losses are taken into account. Conditions for a binary to accelerate very high energy particles, radiate and absorb the non-thermal photons that may reach the observer are discussed. We model the anisotropic structure of the zone of interaction of the relativistic pulsar wind with the strongly magnetized massive star's wind. 
The stellar winds with strong  (in a Gauss range)  magnetic fields at $\sim$ AU distances colliding with powerful pulsar outflows are capable of accelerating particles up to PeV energies at some orbital configurations and phases. The strong magnetic field in the interaction region produces a highly anisotropic structure of the particle accelerator and the emitter in the pulsar outflow. The anisotropic radiation pattern may affect the gamma-ray photon absorption and the number of the observed gamma-ray loud systems.  
\end{abstract}




\end{frontmatter}



\section{Introduction} \label{sec:intro}
Gamma-ray loud binaries (GRLB hereafter) is a very interesting class of gamma-ray sources. It includes by now more than a dozen objects observed in VHE\footnote{\textbf{by convention, we define here the range of {\it high energy} (HE) gamma-rays from 100 MeV to 100 GeV and {\it very high energy} (VHE) gamma-rays -- above 100 GeV}} gamma-rays. There are at least three distinct types of detected GRLBs in which young massive OB-stars have companions of different types. TeV energy gamma-ray emission was detected from $\eta$ Carinae \cite{2025A&A...694A.328H} -- a famous 5.5-years-long period eccentric binary system with colliding winds from a luminous blue variable star and a high-mass companion. Microquasars powered by gas accretion from a normal star onto a stellar-mass black hole (see e.g. \cite{2009IJMPD..18..347B}) were detected as sources of sub-PeV energy gamma-rays \cite{LHAASO_microquasars}. Relativistic winds from young pulsars colliding with wind outflows from OB-stars comprise another type of observed GRLBs \cite{Dubus06a,Dubus13}. The latter two types of GRLBs harboring relativistic stars are producing powerful high-speed outflows, thus providing favorable conditions for proton acceleration to the PeV energies ($1 \:\mbox{PeV} = 10^{15} \:\mbox{eV}$). 

The origin of the observed PeV Galactic cosmic rays is still an open issue. Supernova remnants are the most likely sources of the bulk population of Galactic cosmic rays up to PeV energies. The observed gamma-ray spectra of young supernova remnants have cut-off energies of 100 TeV at most, while
the spectra of some microquasars and GRLBs with pulsars may extend well above 100 TeV \cite{LHAASO_microquasars}. Both microquasars \citep[see e.g.][]{2017SSRv..207....5R,2025arXiv250622550C} and isolated black holes \citep{2025ApJ...981L..36K} are considered as potential PeV particle accelerators.

Extended non-thermal nebulae produced by relativistic winds of isolated pulsars are known as TeV radiation sources, including Crab \citep{Crab_PWN_Lhaaso21,Crab_Amato21},  3C58 \citep{3C58_MAGIC14} and a number of other pulsars presented in the TeVCat catalogue \footnote{https://tevcat2.tevcat.org/}. The detected gamma-ray emission from  isolated pulsar wind nebulae (PWNe) of TeV energies and above can be explained as inverse Compton radiation of leptons accelerated up to multi-TeV energies, with a possible exception of PeV-energy photons detected by LHAASO from the Crab nebula, which might be produced by a quasi-monochromatic component of multi-PeV protons if they are present in the pulsar wind \citep{Crab_Amato21}. A similar situation may be typical for the GRLBs, whose TeV range emission is most likely produced by accelerated leptons, and PeV photons and neutrino can be the result of efficient acceleration of nuclei \citep{2032our} naturally injected into the Fermi acceleration process from the wind collision region. 

Particle acceleration to TeV-PeV energies 
is available even in the case of a pulsar with modest spin-down luminosity $\dot{E}$.
According to a general consideration, the maximum energy of particles accelerated in the outflows of an isolated rotation-powered pulsar 
is limited by its magnetospheric potential \citep[e.g.,][]{arons12,Wilhelmi22}
\begin{equation}
E_{max} = e\sqrt{\frac{\dot{E}}{c}} \approx 0.5 \:\mbox{PeV} \: \dot{E}_{35}^{1/2},
\label{eq:magnetospheric-potential}
\end{equation}
where $e$ is the charge of a proton, $c$ is the speed of light, $\dot{E}_{35} = \dot{E} \, / \, 10^{35} \:\mbox{erg}\,\mbox{s}^{-1}$.

In binary systems, where the particles experience the Fermi type I acceleration in the collision zone of the relativistic pulsar wind and the powerful magnetized wind from the massive star, an important role is played by the anisotropy of both outflows.
The possible presence of rotation-powered decretion disks around rapidly rotating Be-stars\footnote{\textbf{Please note that when the high energy emissions from the GRLBs are discussed the Oe- and Be- stars are refereed together as ``Be-stars'', since in this context the existence of a decretion disk, rather than the spectral type of the companion star or the existence of emission lines is of paramount importance.}} \cite{2013A&ARv..21...69R} may affect the acceleration efficiency. The relativistic pulsar wind is known to be a strongly anisotropic outflow \citep{Michel73,Bogovalov99}; moreover, the anisotropy of pulsar astrospheres inside GRLBs may be enhanced by the strong magnetic field of the stellar wind (SW) in the collision zone \citep{Bykov+24a,Bykov+24b}.
A specific consideration of the maximum attainable energy $E_{max}$ taking into account the anisotropy of a relativistic magnetized outflow produced by the central source with a kinetic/magnetic luminosity of ${\dot{E}}$ was provided in \citep{2009JCAP...11..009L,B12}. Assuming the opening angle of the outflow to be $\Omega$ and its velocity $u_{f} = \beta_{f} c = c \left(1-\Gamma_f^{-2}\right)^{1/2}$,
one may obtain:
\begin{equation}
E_{\rm max} \approx 
1.4 \: \mbox{PeV}\:\frac{{
f\left(\beta_{\rm f}\right)}}{{\rm \Gamma_{\rm f}} \Omega}\:
\dot{E}_{35}^{1/2},
\label{eq:E_max_aniso_outflow}
\end{equation}
where the function $f(\beta_{\rm f}) \propto \beta_{\rm f}^{1/2}$ for $\beta_{\rm f}\ll 1$, and $f(\beta_{\rm f}) \sim 1$ for an ultra-relativistic outflow with Lorentz factor ${\rm \Gamma}_{\rm f} \gg 1$. 

According to these estimates, TeV energies are available even for $\dot{E} \sim 10^{33}-10^{34} \:\mbox{erg}\,\mbox{s}^{-1}$, which is smaller than the known spin-down powers of binary pulsars (see Table \ref{table:grlb-spec-types-and-periods}). 
Strong 
anisotropy of the outflows with $\Omega \sim 0.1$ allows reaching even PeV energies for such pulsar luminosities. Moreover, for higher $\dot{E} \sim 10^{35}-10^{36} \:\mbox{erg}\,\mbox{s}^{-1}$ of established binary pulsars, such an anisotropy allows  
production of multi-PeV protons, making the GRLBs potential Galactic sources of multi-PeV cosmic rays, as well as photons and neutrinos with energies well above 100 TeV. To estimate the possible contribution of GRLBs with relativistic companions to the Galactic pool of very high energy cosmic rays, the expected number of these sources  that account for their duty cycles as PeV accelerators
is required.

The observed number of GRLBs is about a dozen, which is much smaller than several hundred Galactic X-ray binaries.
\textbf{While a young neutron star formed in a massive binary gradually spins-down to become a stellar-wind fed accreting source, the expected number of PSR+OB/Be binaries which can be potential GRLB should be much smaller than X-ray binaries because of shorter life time, details of physics of wind-fed accretion (see, e.g., \cite{2019A&A...622A.189E}), etc. In addition, as we will show in the present paper,}
the anisotropy of the flow structure and VHE emission produced by these sources may make them unobservable in GeV-TeV gamma-rays in a wide range of viewing angles. The above estimates are written for the acceleration of protons; in the case of leptons, the radiative losses in the strong magnetic field and intense radiation field of massive star's emission may put additional constraints on both the maximum energies achieved by these particles and the angular distribution of their VHE emission.
Therefore, a large number of binaries capable of producing TeV-PeV particles may be unidentified as sources of very high energy cosmic rays by the observers on Earth.
To estimate the actual expected number of the Galactic binaries that produce PeV nuclei one needs, on the one hand, the model of their birth and evolution in the Galaxy. On the other hand, one should have a model of particle acceleration in these objects based on the realistic simulations of their structure and a model of production of VHE emission by these particles. 

In this paper, we study the statistics of Galactic GRLBs that can accelerate protons to PeV energies and their observable features. We present here the results of population synthesis simulations of the expected number of Galactic massive binary systems containing a young pulsar that may provide the relativistic wind to interact with the fast outflows from a massive OB- or Be-star companion. We discuss factors that may reduce the number of such objects detected in gamma-rays, including their orbital parameters and the anisotropy of the outflows produced by both binary components. We consider the formation mechanisms of the strong anisotropy of very high energy emission in binaries with a strong magnetic field of the massive star's wind and discuss the expected number of GRLBs capable of producing PeV cosmic rays and VHE emission in the Galaxy. 

\begin{table*}[]
    \centering
    \begin{tabular}{lcccc}
    \hline       
    Type & System & Star spectral type & $P_{orb}$, days & $\dot{E}, \:\mbox{erg}\,\mbox{s}^{-1}$\\
    \hline
    pulsar + Be & \textbf{PSR J2032+4127} & \textbf{Be} &  \textbf{17000} & $\bm{1.7 \times 10^{35}}$ \\
    & \textbf{PSR B1259-63} & \textbf{O9.5Ve} & \textbf{1236.72} & $\bm{8.2 \times 10^{35}}$ \\
    & LSI 61 303 \citep{Weng+22} & Be & 26.49 & $\sim 10^{37}$ \citep{Zabalza+11}?\\
    \hline
    \hline
    compact object + Be & HESS J0632+057 & Be & 315.50 & ?\\
    \hline
    \hline
    compact object + O & HESS J1832-093 & O & 82 & ?\\
    & 1FGL J1018.6-5856 & O & 16.58 & ?\\
    & LMC P-3 & O & 10.2  & ?\\
    & LS 5039$^{*}$ & O & 3.9 & ?\\
    & 4FGL J1405.1-6119 & O & ? & ?\\
    \hline
    \hline
    microquasar & SS 433 & A & 13.08 & - \\
    & V4641 Sgr & B9III & 2.8 & -\\
    & GRS 1915+105 & & & - \\
    & MAXI J1820+070 & & & - \\
    \end{tabular}
    \caption{Orbital periods and spectral classes of massive companions of known gamma-ray binaries of different types. For systems harboring pulsars their spin-down luminosities are also given.  Note that hard X-ray pulsations with a period of $\sim 9$ s were reported by \citep{Yoneda+20} from LS 5039 suggesting that the compact object may be a magnetar.} 
    \label{table:grlb-spec-types-and-periods}
\end{table*}
\section{Population synthesis models}\label{sec:orbits-modeling}

The number of GRLBs and prospects for their detections in high-energy gamma-ray surveys was previously estimated by \cite{2017A&A...608A..59D}. Their estimates varied from a dozen to a hundred such systems in the Galaxy. Here we use a different approach based on the direct calculation of the expected Galactic massive binaries after the first supernova explosion in the parent massive binary system yielding a young rapidly rotating neutron star (NS) acting as a pulsar on orbit around a massive OB or Be-star. The total number of such binaries is normalized to the actual Galactic star formation rate of about two solar masses $M_{\odot}$ per year. This approach enables us to directly calculate the expected 
distributions over binary parameters (orbital period, eccentricities, Be-disk inclinations to the orbital plane), pulsar spin-down power $\dot E$, as well as the type of the massive companion (OB or Be-star) depending on the population synthesis assumptions used. These distributions are needed to further calculate physical conditions in the particle acceleration sites in the wind collision zones.

To simulate the evolution of massive binary systems leading to the formation of GRLBs, we use a modified version of the openly available rapid population synthesis code BSE  \citep{2000MNRAS.315..543H,2002MNRAS.329..897H}
appended by a block for calculation of the spin evolution of magnetized NSs. Unless specified, default input parameter 
values of the binary star evolution are taken from \cite{2002MNRAS.329..897H}. 

Our calculations were performed for the Galactic population I stars with solar metallicity (Z=0.02).
\textbf{In this case, in the BSE code the maximum mass of an early-type star producing a NS in the core collapse is about 20 $M_\odot$.}
To evaluate the model number of GRLBs, we assumed \textbf{a constant star formation rate normalized to
the modern star formation rate in the Milky Way  SFR\,$=2 \,M_\odot \,\mbox{yr}^{-1}$ }
\citep{2011AJ....142..197C,2015ApJ...806...96L}.
The
stellar binarity rate was set to 50\%,
(i.e., 2/3 of all stars are binary components).
The initial mass
function (IMF) of the primary components was assumed to follow the Salpeter law
$(dN/dM \propto M^{-2.35})$ in the mass range $ 0.1M_\odot<M<100 M_\odot$. A
flat distribution of the initial mass ratios of the binary components $q = M_2/M_1 \le 1$
in the [0.1,1] range was used. The initial distribution of 
binaries over orbital periods was accepted in the form  
$f(\log \porb) \propto \log \porb^{-0.55}$ \cite{2012Sci...337..444S}. The initial orbital eccentricities of binary systems were chosen randomly from the range $[0,99]$. \textbf{We set the minimum mass of Be-star to be four solar masses. The criterion for a Be-star to form a decretion disc is the critical equatorial rotation of the star.}
\textbf{In principle, the initial spins of massive binary components may be misaligned due to formation conditions in turbulized molecular clouds. We modeled this assumption in our paper \cite{2019MNRAS.483.3288P}.  However, in the present study we assumed initially coaligned spins of ZAMS binary component.}

One of the major uncertainties in the binary star evolution is the common envelope (CE) stage that arises during heavy mass transfer between the binary components (see \cite{2014LRR....17....3P} and references therein). To treat the change of orbital separation between the binary components  in common envelopes, the $\alpha - \lambda$ formalism 
was adopted, which is based on the comparison of the initial orbital energy of the system and the bounding energy of the donor's envelope \citep{1984ApJ...277..355W,1984ApJS...54..335I}. In our models, we use the fiducial value $\alpha_{CE} = 1$.
To avoid the $\lambda$-description of the envelope binding energy, $\Delta E_\mathrm{env}=GM_\mathrm{env}M_\mathrm{core}/(\lambda R_{\ast})$ (where $M_{env}$ is the envelope mass, $M_{core}$ is the mass of the stellar core, $R_{\ast}$ is the stellar radius, $G$ is the gravitational constant), we directly 
calculated $\Delta E_\mathrm{env}$ using the open-access code described in 
\cite{2011ApJ...743...49L}.
Stellar winds of massive
stars and helium stars were treated using formulas from \cite{2001A&A...369..574V}
and \cite{2017A&A...607L...8V} 
respectively.

We assume that in the core-collapse supernova explosions the nascent NS obtains a kick following a Maxwellian distribution with dispersion
$\sigma = 265$\kms\:  \cite{2005MNRAS.360..974H}. 
Some NSs can be formed due to electron captures in O-Ne-Mg degenerate cores of $8.5-8.8 M_\odot$ stars \cite{2018A&A...614A..99S} 
that experienced a previous mass
exchange \cite{1980PASJ...32..303M}.
The NSs formed in the electron-capture supernovae are assigned a low kick of $30$\kms\, (note that this entirely arbitrary value does not influence the results). 
The kick velocity direction can be isotropically distributed 
or confined within a cone with angle
$\theta_0< \pi/2$ 
coaxial with the progenitor’s rotation axis  \cite{2009MNRAS.395.2087K}.

The BSE code was appended by a block for calculation of the spin evolution of magnetized NSs.
The Scenario Machine code was used as a basis (see \cite{1996A&A...310..489L,2009ARep...53..915L} and references therein). This approach
distinguishes several possible evolutionary stages for a rotating magnetized NS in a binary system that are determined by its interaction with the surrounding plasma: the ejector, propeller, accretor, and
georotator stages. The pulsar phenomenon can appear only on the ejector stage. 
The initial dipole magnetic moments $\hat{\mu}$ of NSs were taken to follow a log-normal distribution \cite{2006ApJ...643..332F}:
\begin{equation}
  f(\hat{\mu})\,d\hat{\mu} \propto
  	\exp\left[-\frac{(\log\hat{\mu}-\log\hat{\mu}_0)^2}{\sigma_{\mu}^2}\right]d\hat{\mu}\,.
\label{mu-norm}
\end{equation}
We made computations for $\log\hat{\mu}_0=30.6$ (corresponding to the surface magnetic field $\log B_0 = 12.6$ for a NS radius of $R_{\rm NS}=10$~km), $\sigma_{\mu}=0.55.$ 
Possible decay of the NS magnetic field was ignored because the lifetime of pulsars (i.e., NSs in the ``ejector'' stage) is much shorter than the characteristic timescale of the NS magnetic field Ohmic decay \footnote{\textbf{Note that the NS magnetic field decay may occur in some cases much faster, see e.g. \cite{2024MNRAS.535L..54I}. However, there is a classical counterexample of an old accreting NS in Her X-1 with the standard $\sim 10^{12}$~G magnetic field.}}.

\begin{figure*}
    \begin{minipage}{.99\textwidth}
        \includegraphics[width=\textwidth]{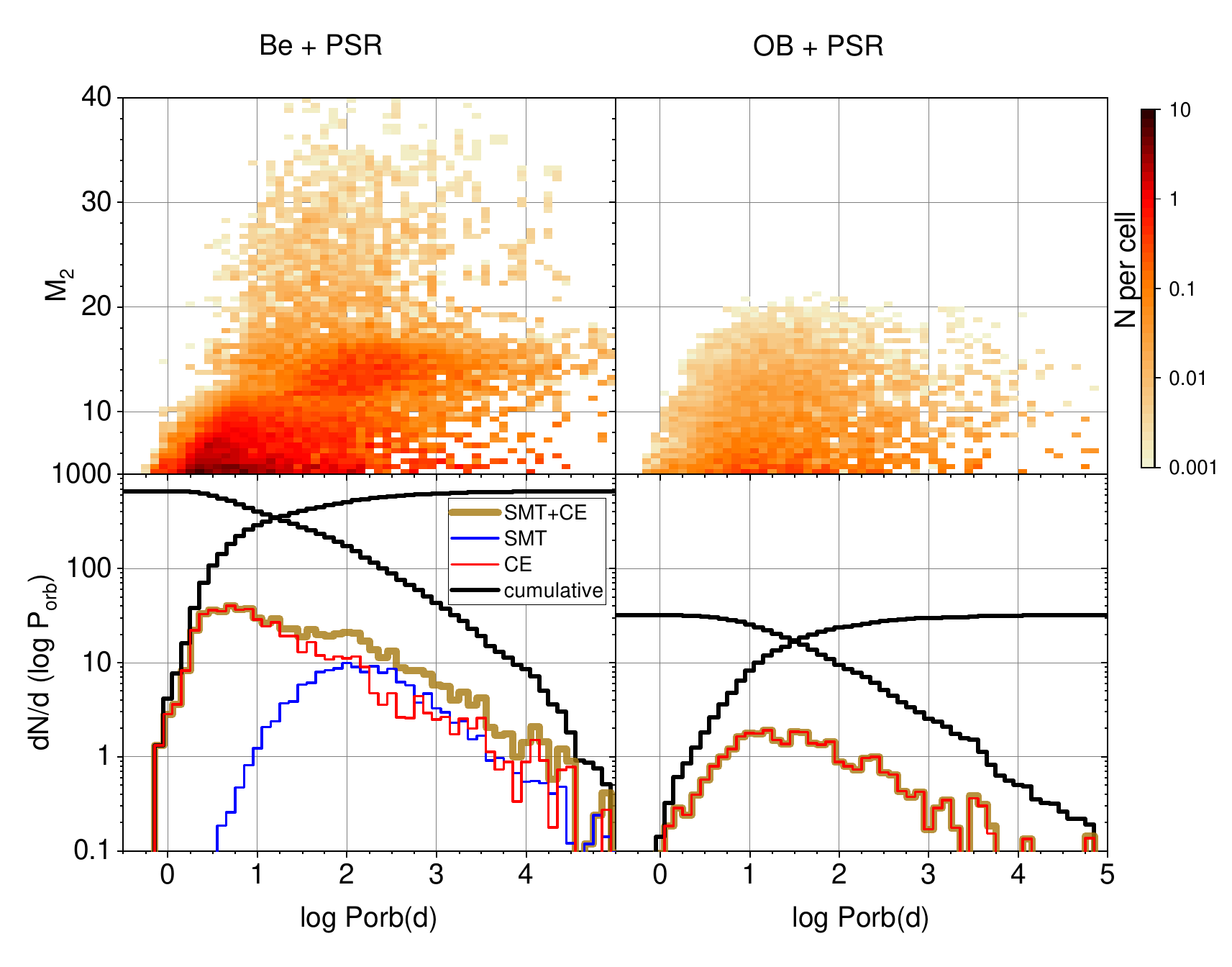} 
   \end{minipage}
   \caption{Simulated Galactic populations of massive binary systems with pulsars for the assumed Galactic star formation rate 2 $M_\odot$/yr. Left and right panels show, respectively, the differential distribution of the Be+PSR and OB+PSR binaries over the optical star mass $M_2$ and the orbital period $P_{orb}$ (top row) and over $P_{orb}$ (bottom row). For Be+PSR systems (left columns), blue and red curves show differential distribution of binaries experienced the first stable mass transfer (SMT) and common envelope (CE), respectively. Brown curve show the total number of systems formed via SMT and CE channels. OB+PSR binaries (right columns) are formed via CE channel only. Black curves show cumulative distributions $N(<\log P_\mathrm{orb})$  and $N(>\log P_\mathrm{orb})$.}
   \label{Be_OB_Nsyst}
\end{figure*}

In the course of massive binary evolution, the more massive primary star first fills its Roche lobe, and the mass exchange between the components starts. In the case of stable episode of the first mass transfer (SMT; see e.g. \cite{Tauris_vdH23}) the secondary component accretes enough mass to become a rapidly rotating Be-star. We used the concept of conservative mass transfer, assuming total mass conservation. In this mass transfer mode the accreting component gains rapid \textbf{(critical)} rotation and keeps \textbf{critically} rotating with increasing mass even after reaching the Keplerian equatorial velocity \cite{1991ApJ...370..597P,1991ApJ...370..604P,1993A&A...274..796B}. After the supernova explosion of the primary component, an eccentric  Be+PSR binary is formed in the range of orbital periods from 10 to 10000 days, with a maximum at about 100 days (Fig. \ref{Be_OB_Nsyst}, blue line in the left panel). 
On the contrary, if during the first mass exchange phase the mass transfer rate from the mass-losing primary star became so high that the secondary companion cannot accommodate all the accreting matter, the common envelope (CE) stage begins. The post-CE binaries (if merging of components did not occur inside the common envelope) become close binary systems in circular orbits. After the supernova explosion of the primary component, orbital periods of such Be+PSR systems fall in a wide range from one to 10000 days, with a maximum at around three days (Fig. \ref{Be_OB_Nsyst}, red line in the left panel).
Note that both stable mass transfer (SMT, about 20\% of systems) and mass transfer ended with the CE stage (about 80\% of cases) can lead to the formation of Be+PSR binaries shown in the left panel of Fig. \ref{Be_OB_Nsyst}.

In principle, OB+PSR binaries in which the OB-star avoided spinning-up during the first mass transfer could result from the evolution of initially very wide massive binaries in which the mass transfer did not occur at all. However, such wide binaries are found to be disrupted during the supernova explosion of the primary component. In our calculations, the progenitors of binary systems with a pulsar and a massive OB-star are found to be binaries with long initial orbital periods and high eccentricities.
In such systems, the time of orbital tidal circularization is longer than the hydrogen burning time of the primary component, and when the primary component's radius increases after the main sequence stage the contact of stars occurs near the orbital pericenter. The contact of the binary components leads to the CE stage. In contrast to the CE occurring after the initial mass transfer episode during Roche lobe overflow stage of circular binaries, we assumed that the CE stage in the eccentric binaries starts so rapidly that the secondary component cannot spin-up during the preceding mass transfer. As a result, after the supernova explosion of the primary, the optical companion of the newborn pulsar will be an OB-type star with a mass near the initial one on the zero-age main sequence (except for possible mass loss by stellar wind).

We stress that in our calculations, OB+PSR systems shown in the right panel of Fig. \ref{Be_OB_Nsyst} are found only after the CE stage occurring shortly after the beginning of the first mass transfer thus preventing the accreting component from gaining rapid rotation to become a Be-star.
\textbf{The SMT mass transfer mode corresponds to the blue curve in the bottom left panel of Figure 1. In contrast, if the first mass transfer is unstable and CE forms, the secondary OB-component does acquire mass in the common envelope and keeps its initial mass $\lesssim 20 M_\odot$ and rotation (OB+PSR systems shown in the right panels in Fig. 1). }.

\section{Statistics of modeled orbits: deficiency of the observed Be+PSR systems}\label{sec:model-statistics-of-GRLBs}

\begin{figure*}
    \begin{minipage}{.99\textwidth}
        \includegraphics[width=\textwidth]{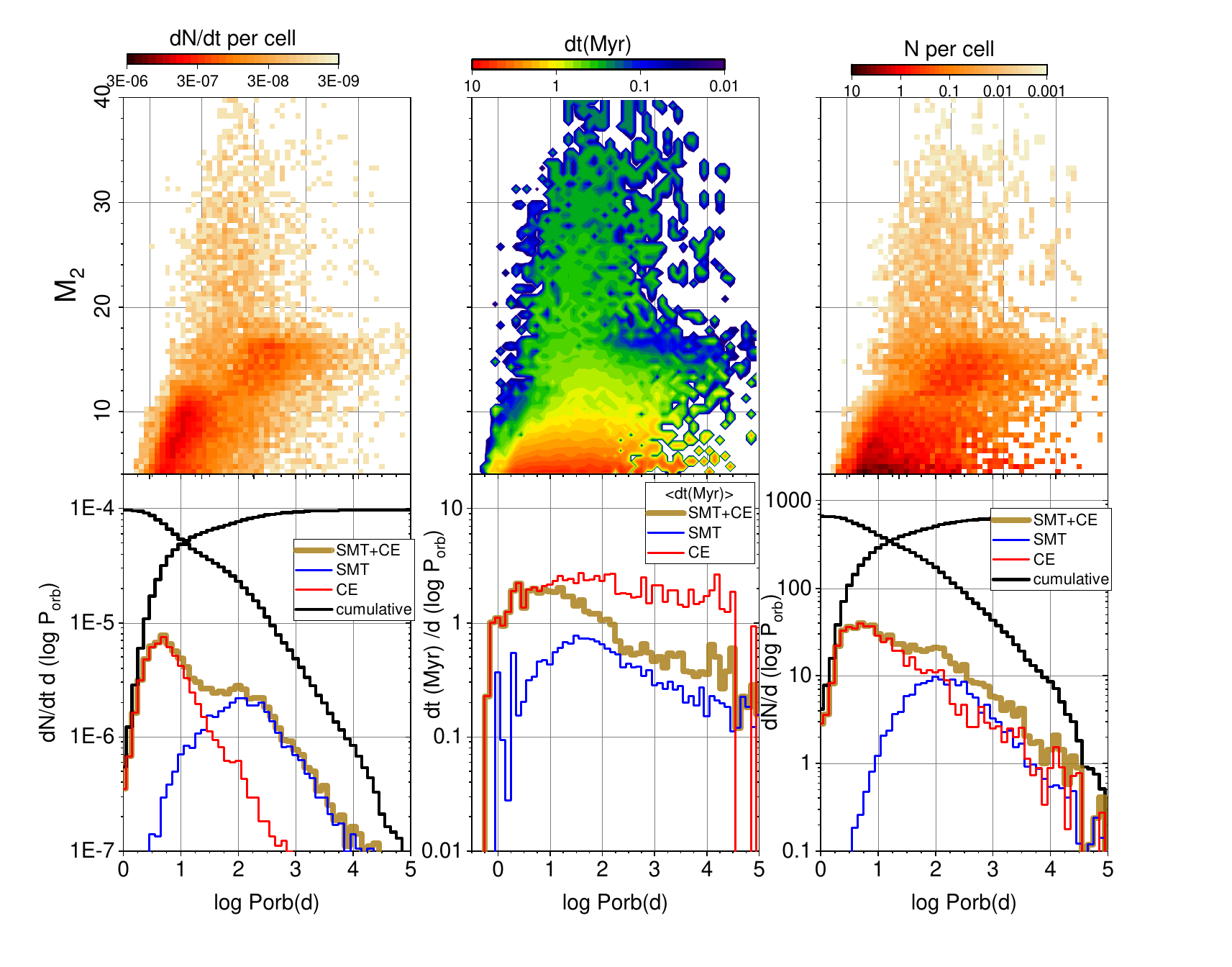} 
   \end{minipage}
   \caption{Simulated population of Galactic pulsar+Be-type star binaries for the neutron star surface magnetic field distributed according to Eq. (\ref{mu-norm}) with $\log B_0 = 12.6$ and $\sigma_\mu=0.55$. Shown are model distributions of Be+PSR systems over the optical companion mass $M_2$ and the orbital period $P_{orb}$ (top row) and over $P_{orb}$ (bottom row). Left panels: distribution of the Galactic formation rate of Be+PSR binaries (per year).
   Middle panels: average lifetime of Be+PSR systems in each parameter bin $\langle dt\rangle$ (in units of $10^6$ years). Right panels: the model distribution of the number of Galactic Be+PSR systems $dN=dN/dt\times \langle dt\rangle$.
   In the  bottom row, blue and red curves show differential distribution of binaries experienced the first stable mass transfer (SMT) and subsequent common envelope (CE), respectively. Brown curve show the sum of the SMT and CE binaries. Black curves show cumulative distributions $N(<\log P_\mathrm{orb})$  and $N(>\log P_\mathrm{orb})$.}
   \label{Be_PSR_dNdtdP_dtdP_dNdP_B_12p6}
\end{figure*}

Our modeling of orbital parameters of Galactic binary systems consisting of a pulsar and a massive Be-type star indicates a strong deficiency in the observed gamma-ray binaries. Let us inspect the bottom-right panel in Fig. \ref{Be_PSR_dNdtdP_dtdP_dNdP_B_12p6} presenting the color-coded model distribution of pulsar+Be star binaries over orbital periods
normalized to the assumed steady Galactic star-formation rate 2~$M_\odot$ per year. In this Figure, the top left panel shows the distribution of the formation rate (per yr) of Be-PSR binaries over the secondary component mass $M_2$ and orbital period. The bottom left panel shows the differential formation rate  of these binaries   (blue, red and brown histograms for SMT without CE, SMT followed by CE formation channels and their sum, respectively) over orbital period $dN/dt/d\log P_{orb}$. The cumulative distribution is in black. The middle panels present the mean time $\langle dt\rangle$ (in Myr) of the model binaries spent in the discrete parameter bins (i.e. the total time all modeled binaries spent in a particular parameter bin divided by the number of modeled systems). The right panels show the simulated number distribution per $M_2$ (top) and per orbital period ($dN/d\log P_{orb}$) of Galactic Be+PSR binaries which can be obtained by multiplication of the left and middle columns $dN=dN/dt\times \langle dt\rangle$.

Binaries with rapidly rotating Be-type massive companions are expected to form in both the SMT and CE scenarios at the pre-supernova stage; the sum of model distributions for these scenarios is shown by the brown curve.
Let us compare the number of such systems predicted by the simulations for different period ranges with the actually observed number of Galactic gamma-ray binaries.
\begin{itemize}
    \item {\it long-period systems, $P_{orb} > 10^4$ d.} The model predicts about one system per period interval at $P_{orb}\sim 10-30$ kilo-days, and the total expected value is $\sim$ 10. Only one gamma-ray binary with such a long orbital period is known: PSR J2032+4127 ($P_{orb} \sim 17000$ d; \citep{Ho+17}).
    \item {\it 1000-day systems}. The model predicts about 30 systems with $10^3 < P_{orb} < 10^4$ d. Only one such system is known -- PSR B1259-63. (Actually, the massive companion LS 2883 is the O9.5Ve star, but Oe and Be stars are generally refereed together as Be-stars, \citep{Chernyakova+24}).
    \item {\it 100-day systems}. About 150 systems with $10^2 < P_{orb} < 10^3$ days are expected according to the model, whereas only one -- HESS J0632+057, $P_{orb} = 317$ d, with yet unknown nature of the compact object -- is observed as the gamma-ray binary \citep{MatchettVanSoelen25}.
    \item {\it 10-day systems}. About 200 Be+PSR systems with $P_{orb} \sim$ few tens days are predicted, only one such GRLB is known -- LSI 61 303, whose compact object is likely a pulsar \citep{Weng+22}.
    \item Finally, the Be+PSR gamma-ray loud {\it short-period systems, $P_{orb} < 10$ d} are unknown, whereas $\sim$ 200 Be+PSR binaries of such period range are predicted by the model.
\end{itemize}

A deficiency of the observed gamma-ray loud binaries in comparison with the model-predicted number of pulsar+Be-star systems is more than obvious. Possible reasons are: (i) either the {\it gamma-ray loud} binaries are rare objects among binary pulsars or (ii) the gamma-ray emission of {\it most} of the pulsar+Be-star binaries is still not detected.

Interestingly, for binaries with OB-stars the deficiency {\it may be} less dramatic, but anyway it is still significant. The model distribution for systems with OB-star is shown in Fig. \ref{Be_OB_Nsyst} (bottom right panel) together with the model for Be-star systems for comparison. According to the model, the binaries with OB-star are an order of magnitude less numerous. 
Just a few systems with $P_{orb} \gsim 10^3$ d are predicted -- and not a single one is observed as a GRLB. For both the short-period systems with $P_{orb} < 10$ d and the systems with $100 < P_{orb} < 1000$ d a dozen of objects is expected, and for the 10-100-day periods about two dozens of binaries are predicted. Among the observed GRLBs with OB-type massive stars one short-period system (LS 5039, $P_{orb} \sim 3.9$ d \citep{Casares05}), two systems with 10-d period (1FGL J1018.6-5856 with $P \sim 16.6$ d, \citep{vanSoelen+22} and LMC P-3  with $P \sim$ 10.3 d, \citep{vanSoelen+19}) and one binary candidate HESS J1832-093 with a longer $P_{orb} \sim 86$ d \cite{Tam+20} are known. 
Assuming that {\it some} of yet unidentified compact objects in these systems are pulsars, one finds that systems with massive OB-stars may be detected as bright gamma-ray sources more often than ones with Be-stars. Indeed, for the 3-4 known Be+PSR GRLBs there are hundreds predicted systems, and for {\it a few} suspected OB+PSR systems then would be just three dozens.    
Why most of the predicted Galactic Be+PSR binaries are not detected as gamma-ray loud objects and what is their difference with gamma-ray loud OB+PSR binaries, if the latter indeed exist, are the issues we discuss in the next sections.


\section{Are the predicted binaries faint? Simple energetics and acceleration efficiency}\label{sec:energetics-and-strong-B}

Let us check whether the pulsar + Be/OB star 
binaries can be absolutely gamma-quiet, i.e. either they do not emit gamma-photons in any direction, or are very faint sources. The gamma-ray emission can be produced by accelerated leptons in the inverse Compton scattering of the optical or ultraviolet (UV) photons of the massive star's radiation or background microwave/infrared/optical photons, including the cosmic microwave background (CMB). Alternatively, the hadrons (protons and nuclei) injected in the Fermi-type acceleration in 
the zone of the binary's winds collision
may gain high energies and emit gamma-photons in p-p or p-$\gamma$ interaction with hadrons of the massive star's wind and photons of stellar emission.

\begin{figure*}
    \begin{minipage}{1\textwidth}
    \begin{minipage}{.8\textwidth}
        \includegraphics[width=\textwidth]{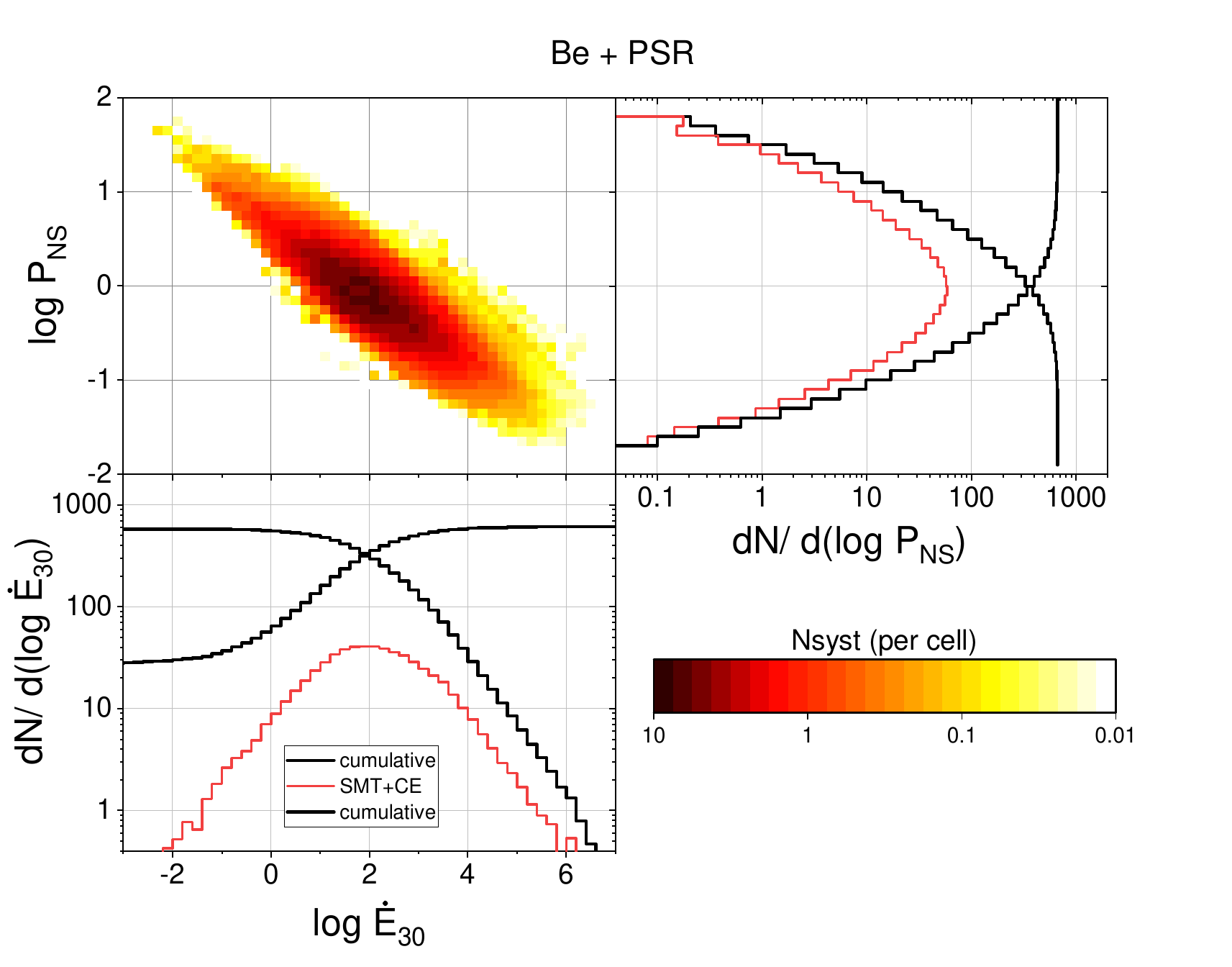} 
   \end{minipage}
   \linebreak
   \vfill
   \begin{minipage}{.8\textwidth}
        \includegraphics[width=\textwidth]{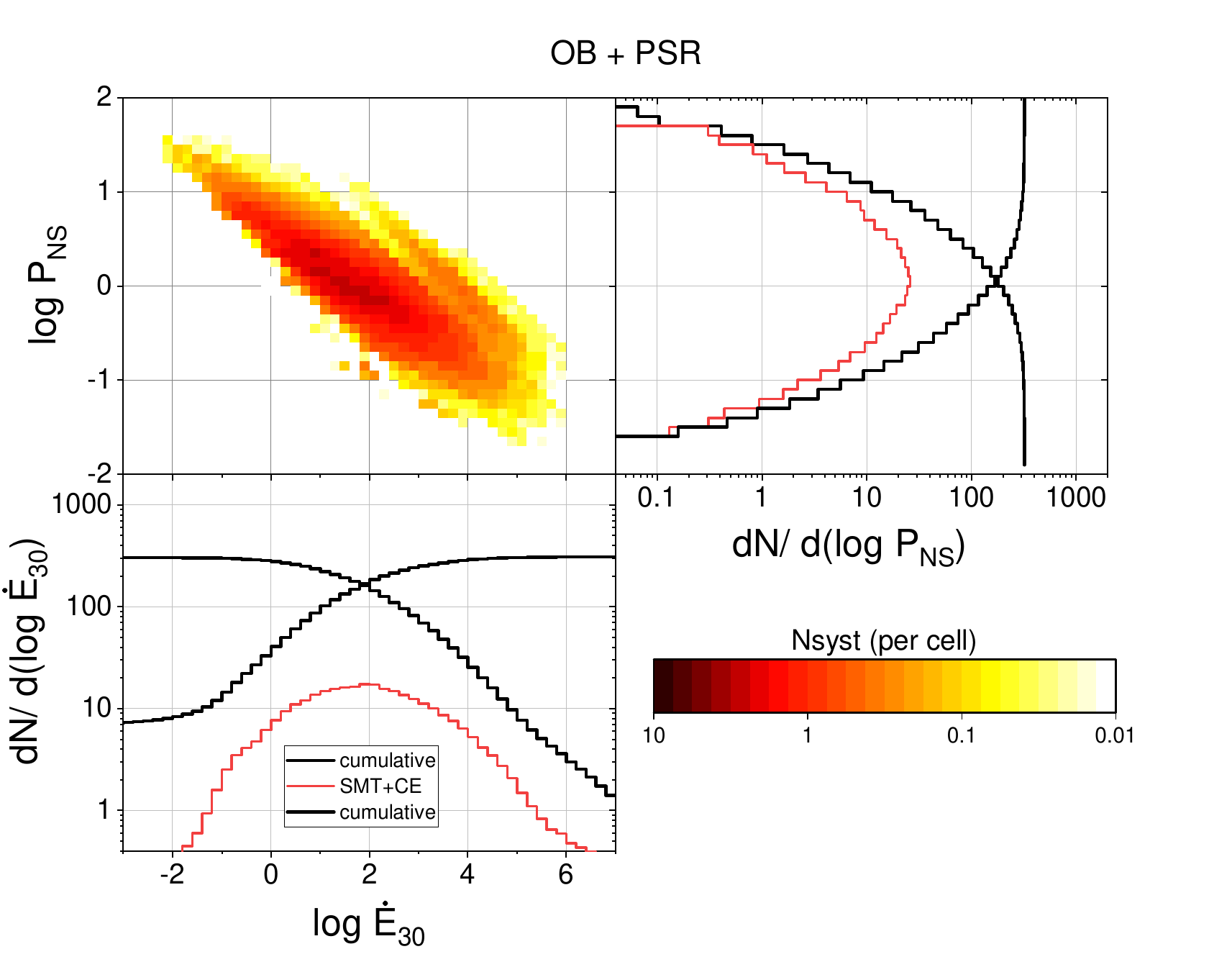} 
   \end{minipage}
   \caption{Model distribution of Galactic number of binary systems with pulsars over the spin-down power of pulsars (in units $\dot E_{30} =10^{30}\ergs$) and the neutron star spin period $P_{NS}$ for Be+PSR (top panels) and OB+PSR (bottom panels) binaries.
   }
\label{fig:Be-OB-distr}
\end{minipage}
\end{figure*}

A low spin-down power of the pulsars in the predicted hundreds of binaries can prevent their detection by the modern gamma-ray observatories. Two established pulsars in the GRLBs have spin-down luminosities of order $10^{35}-10^{36} \:\mbox{erg}\,\mbox{s}^{-1}$. The model distributions of the PSR+Be and PSR+OB binaries over $\dot{E}$ are given in Figure \ref{fig:Be-OB-distr}. The model predicts about a dozen systems with $\dot{E} \gsim 10^{35} \:\mbox{erg}\,\mbox{s}^{-1}$ for both types of massive components, that is twice as numerous as the known population of GRLBs.
However, large distances to some of these objects can make them too faint. A simple estimate of the expected VHE gamma-ray flux from the binary system that is located at a distance $d$ and converts a fraction $\eta$ of the pulsar's $\dot{E}$ in the VHE gamma-rays {\it in case of isotropic energy release} can be written as follows:
\begin{equation}
F_{VHE} = \frac{\eta \dot{E}}{4 \pi d^2} = 4 \times 10^{-14} \eta_{-2} d_{15}^{-2} \dot{E}_{35} \:\:\mbox{erg}\,\mbox{cm}^{-2}\,\mbox{s}^{-1},
\label{eq:VHE-flux-with-distance}
\end{equation}
where $\eta_{-2} = \eta / 10^{-2}$, and $d_{15} = d / \left(15 \:\mbox{kpc}\right)$. If some of the expected binaries with a high spin-down power are located in distant regions of the Galaxy, they can be detected just with a very long exposure.

The estimate given by Eq.(\ref{eq:VHE-flux-with-distance}) does not take into account a number of factors that can strongly influence the VHE flux produced by the binaries, either increasing or decreasing the probability of their detection in the VHE range. In the following we will discuss in detail some of these factors. Let us start from the acceleration efficiency. 

The binary systems with rotation-powered pulsars, where the relativistic wind of a pulsar collides with the powerful wind from a young massive star, provide favorable conditions for efficient particle energization. In these systems, Fermi type I acceleration in the colliding flows may be at work \citep{Falanga+21,2032our}. In this process hard particle spectra $f\left(E\right) \propto E^{-s}$, $s<2$ are produced in the energy interval $E_{min}^{cwf} < E < E_{max}^{cwf}$, where the injection energy $E_{min}^{cwf}$ is the minimum energy allowing multiple crossings of the contact discontinuity between two winds by a  particle, and $E_{max}^{cwf}$ is either the maximum energy allowing particle confinement in the colliding winds zone or the energy above which the radiative energy losses of the confined particles become more rapid than the particle acceleration. This produces a growing spectral energy distribution for $E_{min}^{cwf} < E < E_{max}^{cwf}$ and makes possible the conversion of a significant fraction of the pulsar spin-down power into the energy of particles with $E \sim E_{max}^{cwf}$. 

High values of $\eta > 10^{-2}$ may be required to explain with GRLB models some very powerful (sub)PeV flares from the Cygnus region as that reported by \cite{Carpet2}.
To reach PeV particle acceleration with very high efficiency of the source kinetic luminosity conversion to PeV regime particle, one needs a very efficient confinement of particles at the acceleration site provided by strong CR driven turbulence there, as it was illustrated with a Monte Carlo modeling \cite{2032our}.

\section{Strong magnetic field in the collision region}\label{sec:strong-B-and-radiative-losses}
\begin{figure}
\center{\includegraphics[width=\columnwidth]{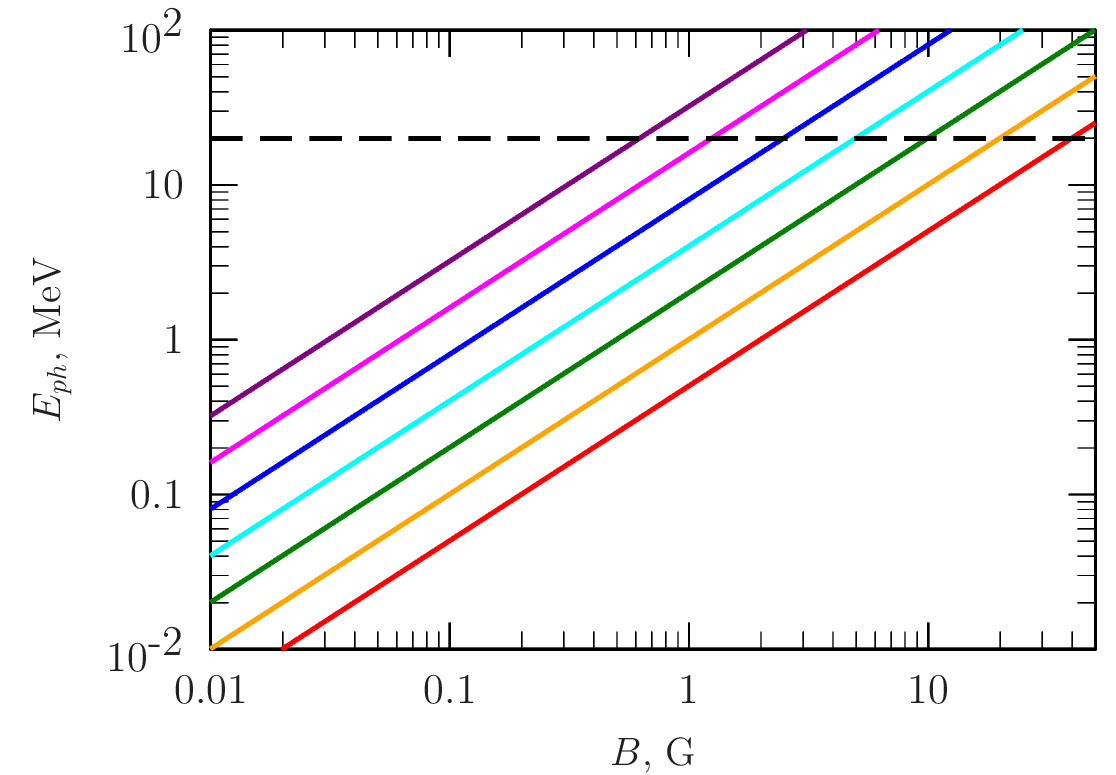}}
\caption{\label{fig:LS-5039-max-E}
Estimate of the magnetic field amplitude in the gamma-ray binary LS 5039. 
Color curves show a typical energy of synchrotron photons produced by electrons of different energies $E_{max}^{cwf}$ (fixed for each curve) as a function of the magnetic field in the emission site $B$. $E_{max}^{cwf}$ is the maximum energy available in particle acceleration in colliding winds of the binary.
Different colors correspond to different values of $E_{max}^{cwf}$, with $\gamma_{max} = E_{max}^{cwf} / mc^2$, chosen logarithmically uniform from $10^7$ to $8 \times 10^7$ (from red to purple). This range covers the expected maximum energies of power-law component in particle spectra capable of explaining the TeV spectrum and its cut-off energy for LS 5039 in the inferior conjunction. The TeV photons can be produced in the inverse Compton process by the same electrons that emit synchrotron keV-MeV radiation \citep{Falanga+21}. The dotted line shows the energy of the spectral maximum in LS 5039  ($E_{ph}=$ 20 MeV), which is likely to be the maximum of the synchrotron emission component. The intersection of a color curve with the dotted curve yields the estimation of the magnetic field needed for the model consistency: e.g., assuming $\gamma_{max} = 4 \times 10^{7}$ (blue curve), one gets $B \sim 3 \:\mbox{G}$.}
\end{figure}

A strong Gauss-range magnetic field is required to explain a number of observational features of known GRLBs.
\begin{itemize}
    \item Matching of maximum energy of accelerated particles $E_{max}^{cwf}$ inferred from the synchrotron (keV-MeV) and the inverse Compton (TeV) spectral components of LS 5039 implies a Gauss-range field (Fig. \ref{fig:LS-5039-max-E}).
    \item Photons and neutrino $E > 100 \:\mbox{TeV}$ associated with a number of GRLBs assume proton acceleration above PeV in these systems, which requires Gauss-range magnetic field. Such a field allows to accelerate protons well above PeV without strong cosmic ray driven turbulence \cite{Bykov+24a}, while the explanation of extremely strong flares in Cygnus region reported in \cite{Carpet2} requires extreme conditions of particle scattering in the winds collision region \cite{2032our}.
    \item To explain the variability of radio emission of PSR B1259-63, a very strong magnetic field $\sim$ 20 G may be required \citep{Chernyakova+24}.
\end{itemize}

The Gauss-range magnetic field in the winds collision region can be typical for the pulsar+massive star binaries as it is readily provided by the stellar wind.
Indeed, typical orbital separations in GRLBs lie in the range $R \sim 10^{12} - 10^{14}$ cm.
Then, assuming Parker's model and typical massive star's radius $R_{\ast} \sim 10 R_{\odot}$, the estimated value of the SW magnetic field at distance $R$ from the massive star $B_{sw}  = B_{\ast} R_{\ast}/ R$ is $\sim 1$ G for the massive star's surface magnetic field $B_{\ast} \sim 100 \:\mbox{G}$ for any $R$ of interest. It still falls into the Gauss range for much lower $B_{\ast} \sim 1 - 10 \:\mbox{G}$ for sub-AU orbital separations.
Studies of the massive star's surface magnetic field confirm that the required amplitudes of $B_{\ast}$ can be expected; moreover, the
dipolar fields in about 10\% of O/B/A stars are found to exceed 100 G \citep{OBstars_Bfield17,Bstar_Bfield19}. 
Note, for typical parameters of stellar winds, such magnetic field amplitude may imply strong magnetization of the wind.

Since the VHE emission of observed GRLBs is most likely associated with the inverse Compton radiation from leptons, which are a subject of radiative energy losses, one needs to check whether the Gauss-range magnetic field is compatible with the TeV-range leptons production. 
The maximum energy that may be reached by leptons in GRLBs can be estimated by matching the timescales of particle acceleration and radiation losses.
Synchrotron loss rate by a lepton with mass $m$ and energy $E$ in a uniform magnetic field with amplitude $B$ is given by Eq.  (\ref{eq:synchro_rate}):
\begin{equation}
-\frac{dE}{dt} = \frac{2}{3}\frac{e^4}{m^2c^3}B_{\perp}^2\left(\frac{E}{mc^2}\right)^2.
\label{eq:synchro_rate}
\end{equation}
This rate depends on $B_{\perp} = B \sin\chi$ -- the component of the magnetic field {\it transverse} to the particle velocity direction, where $\chi$ is the pitch angle between the particle velocity and magnetic field. For leptons with energy in TeV range with $\gamma = E / mc^2 \sim 10^6-10^7$ the inverse Compton radiation losses are produced by upscattering of the CMB photons and photons of optical/UV emission of the massive star (in the Klein-Nishina limit).

Leptons gain TeV energies inside the PWN bubble inflated by the pulsar wind in the collision winds zone: at the pulsar wind's termination shock and in the colliding flows of the shocked pulsar wind and the SW.
The estimate of $\tau_a$ -- the typical acceleration time of a particle from the initial energy $E_0 = mc^2\gamma_0$ up to $E = mc^2\gamma$ in the collision zone of the relativistic wind with an upcoming flow -- was given in \citep{BSPWN_2017}:
\begin{equation}
\tau_a \approx \frac{9}{c}\int\limits_{\gamma_0}^{\gamma} \frac{D\left(\gamma\right)}{u}\frac{d\gamma}{\gamma}
\label{eq:acceleration-time-in-cwf}
\end{equation}
In Eq. (\ref{eq:acceleration-time-in-cwf}) $u$ is the upcoming flow velocity, $D$ is the diffusion coefficient in the flow. Assuming the  Bohm diffusion with $D = cR_g/3$, where $R_g = E /eB$ is the particle gyroradius in the magnetic field  $B$, one obtains
\begin{equation}
\tau_a \sim \frac{3R_g}{u} \approx 100 \:B_{sw}^{-1}\left(\frac{E}{1 \:\mbox{TeV}}\right)\left(\frac{u_{sw}}{10^3 \:\mbox{km}\,\mbox{s}^{-1}}\right)^{-1}, \:\mbox{s}
\label{eq:acceleration-time-numbers}
\end{equation}
The timescale of the synchrotron losses dominating at considered energies for $B \sim$ a few Gauss at typical orbital separation distance from the massive star (see the comparison with the inverse Compton losses timescale in Fig. \ref{fig:t_cooling_leptons}) is given by Eq. (\ref{eq:synchro_time}):
\begin{equation}
t_{syn} = \frac{3}{2}\frac{m^3c^5}{e^4}\frac{1}{\gamma B^2\sin^2\chi} \approx \frac{3 \times 10^2}{ B^2\sin^2\chi}\left(\frac{E}{1 \:\mbox{TeV}}\right)^{-1},\:\mbox{s}
\label{eq:synchro_time}
\end{equation}
The comparison of Eq. (\ref{eq:acceleration-time-numbers}) and (\ref{eq:synchro_time}) shows that in a binary system with large $B_{sw} \sim $  a few G leptons can be accelerated up to TeV energies. Indeed, taking the averaged value of $B^2_{\perp} = (2/3) B^2$ and $B = B_{sw} = 4 \:\mbox{G}$, one gets $t_{syn}\gsim\tau_a$ for $E = 1 \:\mbox{TeV}$.

Thus, according to simple estimates of attainable particle energy, 
binaries with Gauss-range magnetic field in the collision winds region seem to be able to produce gamma-ray photons by leptonic mechanism at least in GeV-TeV range. In fact, modeling of the structure of magnetized flows in GRLBs we discuss below shows that in a significant part of the accelerating volume $B$ is below the $B_{sw}$ value and lies in the sub-Gauss range, so the synchrotron losses can be slower. A sophisticated modeling shows that systems with $B_{sw}$ in the Gauss range can accelerate leptons up to even larger energies $\gamma \sim 10^7$ allowing us to explain the observed MeV and TeV radiation in LS 5039. 

According to our findings, (i) binaries with colliding winds may have strong Gauss-range magnetic fields that help to explain their high-energy observations; (ii) this field does not prevent them from converting a significant fraction of the pulsar's spin-down power into gamma-rays, at least in the GeV-TeV range.
The influence of Gauss-range fields as well as of the strong SW magnetization on the observational properties of GRLBs is still poorly investigated. In the following, we draw a number of important conclusions assuming these conditions. 

\section{Orbital geometry and Be-disk inclination}\label{sec:geometry-and-inclination}

The need of a strong magnetic field to  produce high-energy radiation can result in variability of these emissions with orbital period of a gamma-ray binary. Indeed, one should not expect that the physical conditions, including the SW magnetic field amplitude, would be uniform along the entire extended orbit of the pulsar. For example, even a small eccentricity of the orbit may result in different spectra in the TeV range.

Short-period binaries with small orbital semi-axes of the order of a few tenths of AU highly likely can produce high energy emissions over the whole orbital period. A prominent example is LS 5039 with $P_{orb} = 3.9$ d and eccentricity $e \sim 0.35$ \citep{Casares05} observed by H.E.S.S. both in the inferior and superior conjunction orbital phases \citep{Aharonian06}.
At the same time, the TeV-range spectral energy distribution in these phases dramatically differs, with a flat distribution in the inferior conjunction and a rapid fall in the superior one.
More eccentric orbits of known gamma-ray binaries may alternate magnetic field amplitude, radiation field intensity, and other parameters governing the particle acceleration and the VHE emission production and absorption along the orbit even much stronger than in LS 5039.   

Powerful winds produced by young massive stars are known to be anisotropic, with a dense magnetized equatorial component and rarefied outflows at higher latitudes \cite{2022ARA&A..60..203V}. The most pronounced anisotropy is found for rapidly rotating Be-type stars with their equatorial disks. 
Passage of the pulsar through the magnetized dense equatorial disk provides conditions for more efficient particle acceleration in the colliding flows because the strong magnetic field supports particle confinement in the acceleration zone. Thus, the conditions for the production of high energy emissions may be very different along the orbit around a Be-star, if (i) the orbit is eccentric and the pulsar passes through the disk only during a part of the orbital period, or (ii) the disk is inclined with respect to the orbital plane.

\begin{figure*}
    \begin{minipage}{.99\textwidth}
        \includegraphics[width=\textwidth]{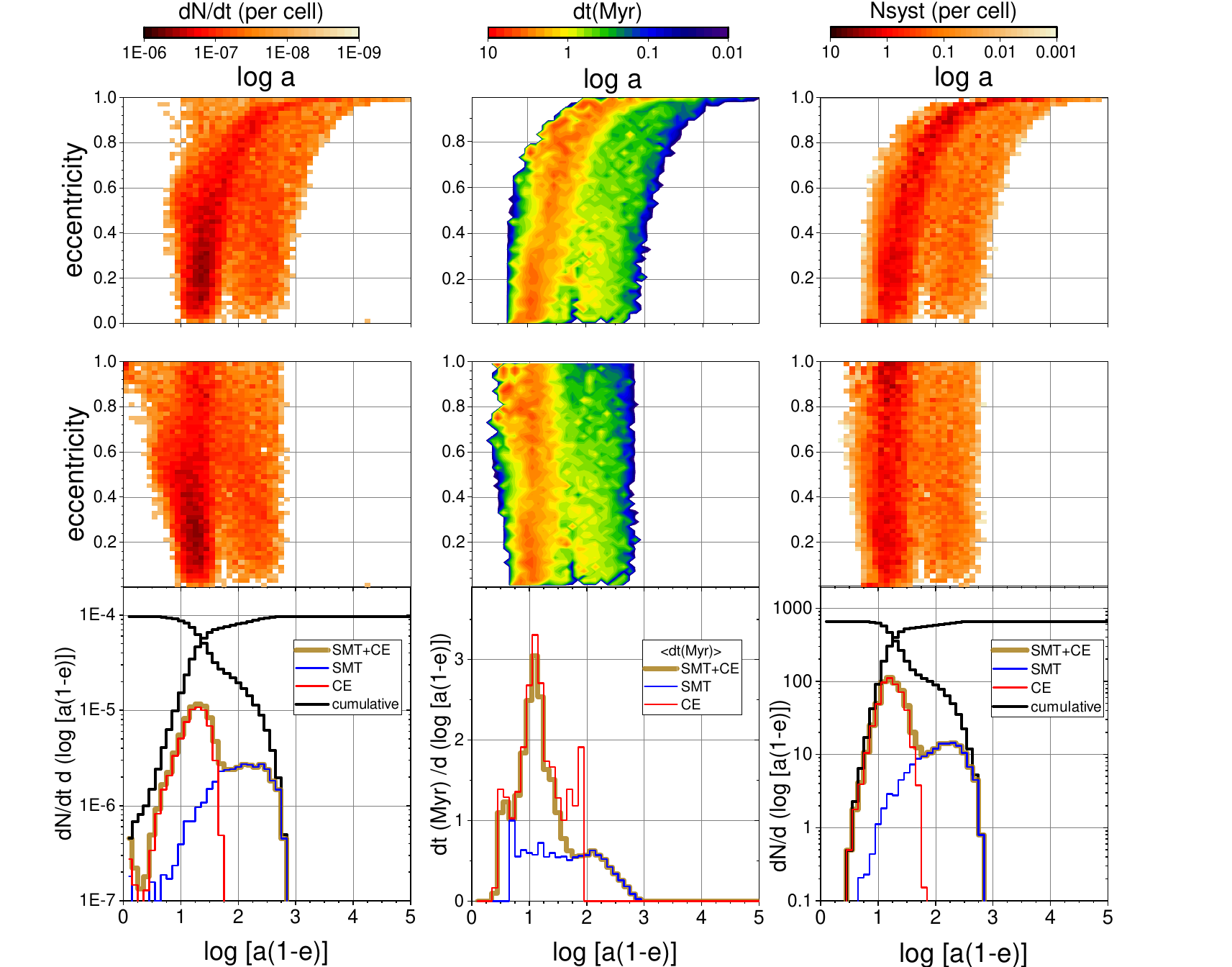} 
   \end{minipage}
   \caption{Model distributions of Galactic Be+PSR binaries 
   over eccentricity $e$ and orbital periastron distance $R_{min}=a\left(1-e\right)$ ($a$ is the orbital semi-major axis in units of solar radius). As in Fig.2, the left columns show the distribution of Be+PSR formation rate (per year), the middle columns show the average time the systems spent in the particular parameter bin (in Myrs), and the right columns is the number distribution per parameter bin, i.e. the product of the left and middle distributions. Top row: 2D model  distributions over $e$ and $a$. Middle row: 2D model  distributions over $e$ and $R_{min}$. Bottom row -- 1D distributions over $R_{min}$. The calculations are shown for the neutron star surface magnetic field
   parameter $\log B_0 = 12.6$.}
   \label{fig:Be-PSR-model-ecc}
\end{figure*}

\begin{figure*}
    \begin{minipage}{1\textwidth}
    \begin{minipage}{.78\textwidth}
        \includegraphics[width=\textwidth]{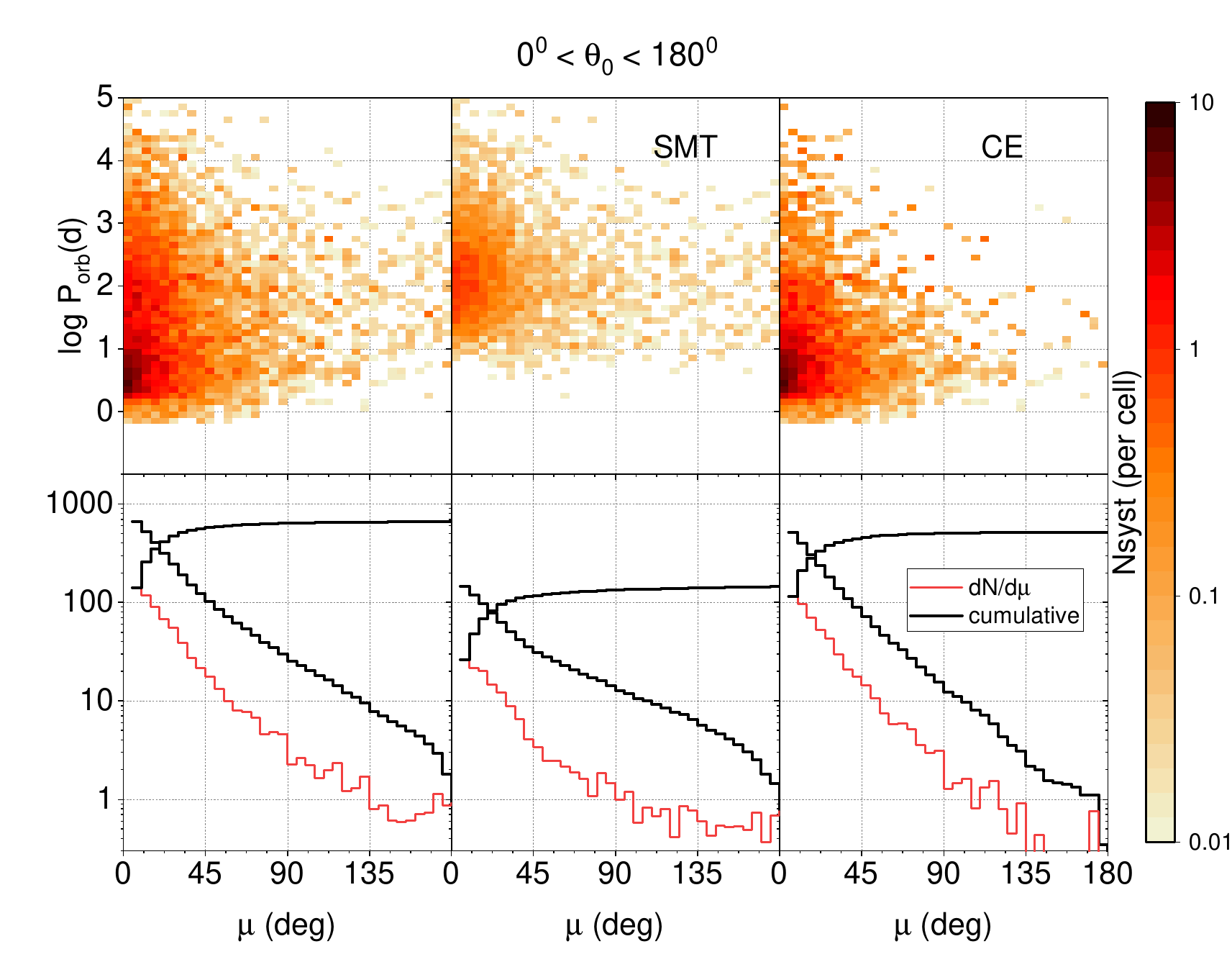} 
   \end{minipage}
   \linebreak
   \vfill
   \begin{minipage}{.78\textwidth}
        \includegraphics[width=\textwidth]{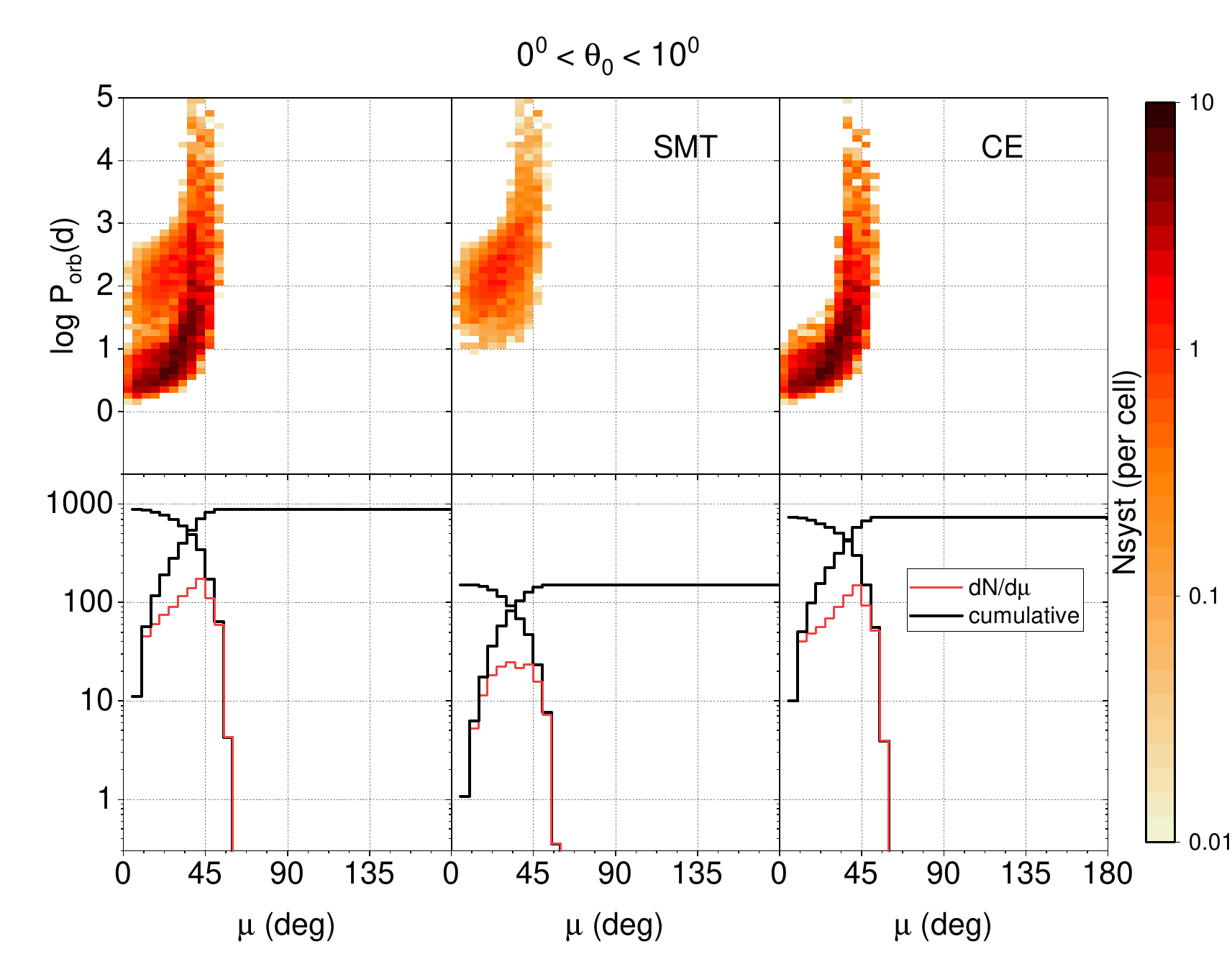} 
   \end{minipage}
   \caption{Model distributions of pulsar+Be-star binaries over angle $\mu$ between the spin axis of the massive star and the orbital angular momentum. Top panel: the kick velocity produced in the parent supernova explosion is distributed isotropically for $0^{\circ}<\theta_0<180^{\circ}$, where $\theta_0$ is measured from the pre-supernova spin axis. Bottom panel: the kick is isotropic within a narrow cone with $0^{\circ}<\theta_0<10^{\circ}$. 
   The distributions for SMT scenario and for scenario with common envelope (CE) are in the middle and right panels respectively. Left panels -- sum of SMT and CE distributions.
   }
\label{fig:Be-PSR-model-mu-distr}
\end{minipage}
\end{figure*}

Modeling of pulsar+Be-star binaries showed that the orbits 
of $\sim$20\% of these systems that were produced after the stable mass transfer (see Sec. \ref{sec:orbits-modeling}) have
relatively large periastron distances $\gsim 10^2 R_{\odot}$ (see bottom panels in Fig. \ref{fig:Be-PSR-model-ecc}). In turn, the large periastron distances seem to be associated with a high orbital eccentricity, $e\rightarrow 1$. Interestingly, PSR B1259-63 and PSR J2032+4127 do have eccentric orbits. The modeling also has shown that the orientation of the Be-star disk with respect to the pulsar's orbit may be different. In Fig. \ref{fig:Be-PSR-model-mu-distr} we show model distributions of such systems over angles between the Be-star spin axis and the system's orbital spin axis $\mu$, as well as over the orbital period $P_{orb}$. In the case of isotropic kick after the parent supernova explosion (upper panel of Fig. \ref{fig:Be-PSR-model-mu-distr}) the distribution is concentrated at small angles $\mu\sim 0^\circ$ (especially in binaries formed after the CE stage, the right columns), i.e. the disk typically lies in the binary orbital plane. However, for short and medium orbital periods about a percent of binaries can have the disk perpendicular to the orbital plane ($\mu=90^\circ$). In the case of a strongly anisotropic kick (lower panel of Fig. \ref{fig:Be-PSR-model-mu-distr}), systems with medium and long orbital periods would have disks tilted at angles reaching $\sim$ 45$^{\circ}$.

In the binary system of PSR B1259-63 the disk of the Be-type massive companion LS 2883 is tilted to the orbital plane by $\sim$ 35$^{\circ}$ \citep{Shannon+14}. Due to that, at the periastron passage the system's non-thermal emission shows a non-trivial variability associated with efficient acceleration of particles at two crossings of the disk before and after the periastron and eclipse of the pulsar in-between, as well as a change of the orientation of the winds collision region \citep{Chernyakova+09,Chernyakova+14,Chernyakova+24,Chernyakova+25,HESS_B1259_2024}. The binary system PSR J2032+4127/MT91 213 may also harbor an inclined Be-disk. Such an assumption allows one to explain the nature of a bright sub-PeV photon and neutrino flare detected in 2020 simultaneously by {\sl Carpet-2} and {\sl IceCube} from the Cygnus region and being consistent spatially with the position of PSR J2032+4127 \citep{Carpet2}. This flare under extreme conditions may be produced by rapid efficient acceleration of protons above 10 PeV and rapid emission of their energy in the photomeson process during the passage of the pulsar through the strongly magnetized Be-star's equatorial disk significantly tilted to the orbital plane \citep{2032our}. On the other hand, this flare may be produced by the very powerful outflow from  the microquasar Cyg X-3 located nearby.

Both possibilities of high orbital eccentricity  and high inclination of the Be disks can result in strong variability of high energy emissions of considered pulsar+Be star binaries. Their GeV-TeV and even PeV emission may manifest as periodic flares, which may be almost elusive for distant observers unless such behavior of the source is already known. These factors of orbital orientation and geometry may reduce the number of the binary systems observed as GRLBs. Moreover, pulsar eclipse by the dense equatorial disk may enhance the $\gamma$-$\gamma$ absorption of the high energy emission produced at the periastron and further reduce the probability of detection of these objects in gamma-rays.

\section{Strong magnetic field and anisotropic pulsar astrospheres}\label{sec:anisotropic-PWNe}

Recently, Bykov et al. \citep{Bykov+24a,Bykov+24b} performed relativistic magnetohydrodynamical (rMHD) simulations of the structure of magnetized flows in the winds collision zone in a binary with a strongly magnetized wind of the massive companion\footnote{For consideration of the anisotropic cold pulsar wind interacting with the non-magnetized stellar wind see \citep{bosch-ramon21} and the references therein.}. These simulations showed that a strong Gauss-range magnetic field of the stellar wind $B_{sw}$ may provide a peculiar elongated shape of the bubbles of PWNe blown by the relativistic pulsar winds. The strong magnetic field prevents the PWN's expansion across its direction, so the PWN becomes extended along it. This anisotropy becomes more and more pronounced with the growth of $B_{sw}$.  While at sub-Gauss amplitudes a slightly anisotropic shape of the PWN bubble is governed by the latitudinal anisotropy of the pulsar wind, at large $B_{sw} > 1$ G it is already governed by the stellar wind's magnetic field, and for $B_{sw} \sim 3-4$ G the bubble's length becomes several times larger than its width.

  \begin{figure*}
\begin{minipage}{1.0\textwidth}
\begin{minipage}{.43\textwidth}
        \includegraphics[width=\textwidth]{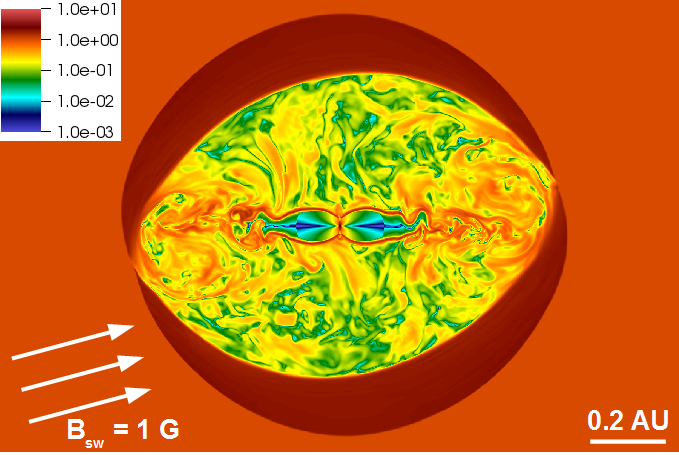}
   \end{minipage}
   \nolinebreak
   \hfill
   \begin{minipage}{.43\textwidth}
        \includegraphics[width=\textwidth]{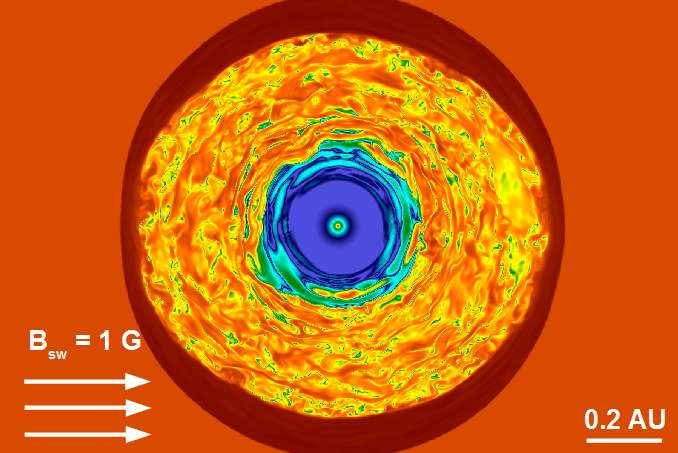}
   \end{minipage}
    \linebreak
   \vfill
   \begin{minipage}{.43\textwidth}
     \includegraphics[width=\textwidth]{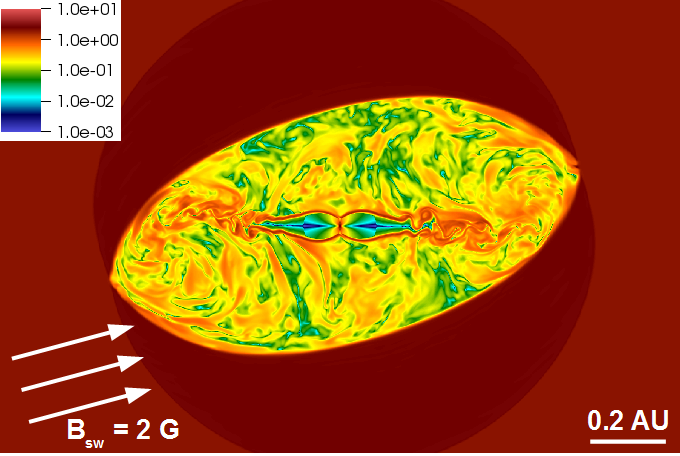}
   \end{minipage}
   \nolinebreak
   \hfill
   \begin{minipage}{.43\textwidth}
     \includegraphics[width=\textwidth]{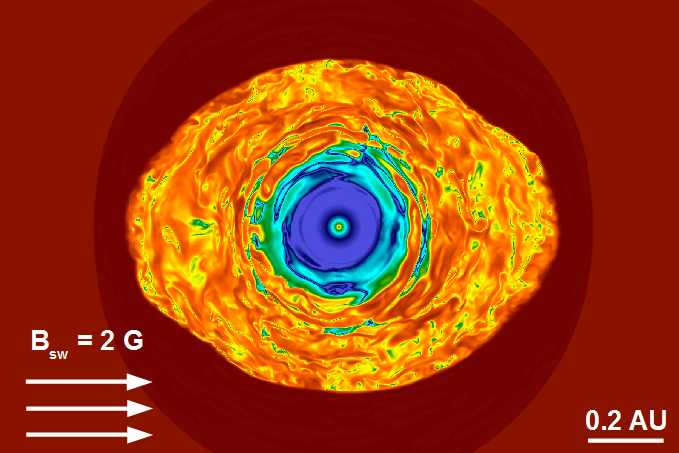}
   \end{minipage}
   \linebreak
   \vfill
   \begin{minipage}{.43\textwidth}
     \includegraphics[width=\textwidth]{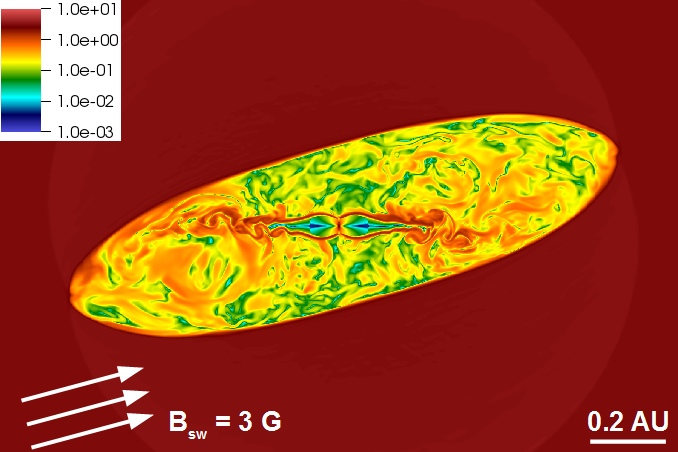}
   \end{minipage}
   \nolinebreak
   \hfill
   \begin{minipage}{.43\textwidth}
     \includegraphics[width=\textwidth]{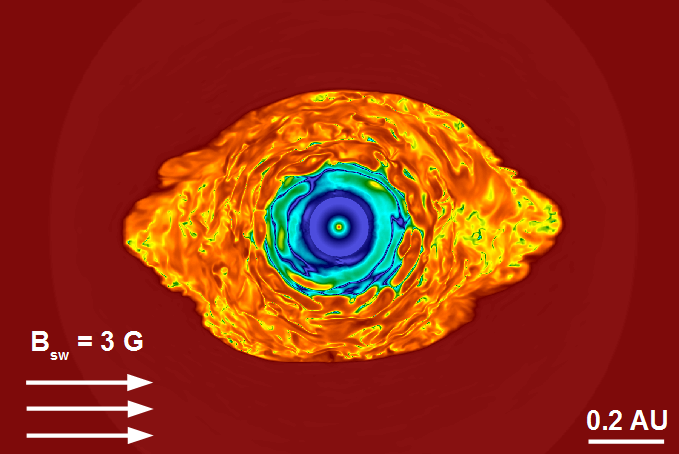}
   \end{minipage}
    \linebreak
   \vfill
   \begin{minipage}{.43\textwidth}
     \includegraphics[width=\textwidth]{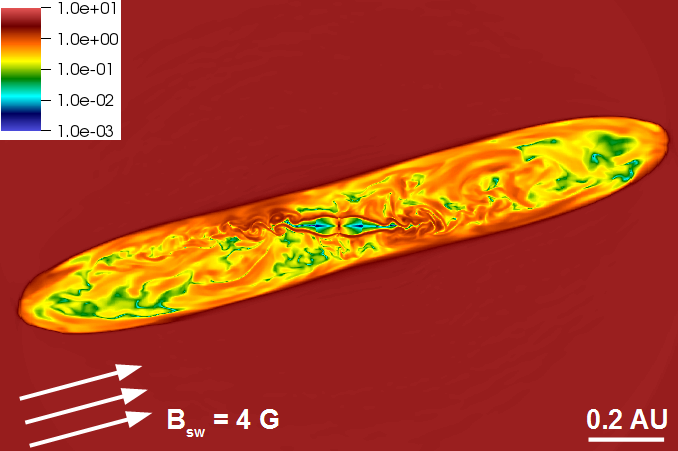}
   \end{minipage}
   \nolinebreak
   \hfill
   \begin{minipage}{.43\textwidth}
     \includegraphics[width=\textwidth]{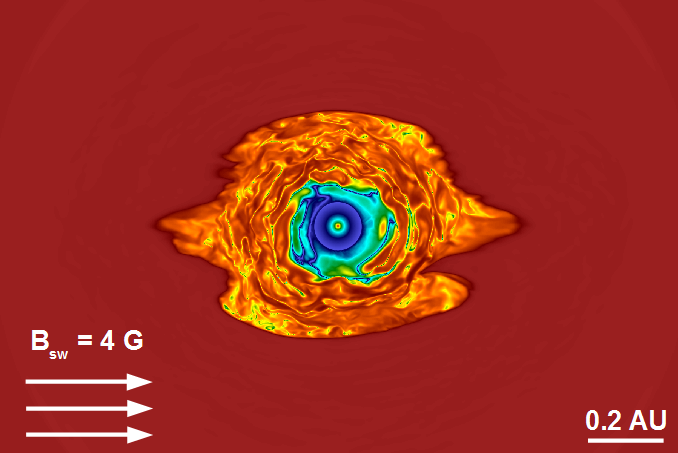}
   \end{minipage}
 \caption{\label{fig:rMHD-PWN-vs-B}
 Relativistic MHD simulations of the collision zone of two winds in a gamma-ray binary (performed using code PLUTO \citep{Mignone+07}). Shown is the structure of the magnetic field $B$ (in Gauss) produced in modeling of the pulsar wind nebula inflation in a strongly magnetized locally uniform stellar wind flow. The results of 3D simulation are shown in two cuts: meridional (left columns) and equatorial (right columns). The latter is the plane of pulsar's rotational equator, the former is defined by two vectors belonging to it: the direction of the pulsar's spin axis (directed upwards in left panels) and the direction of the local large-scale magnetic field of the stellar wind (shown by white arrows in both columns). Rows show the results of simulation for different amplitudes of the stellar wind magnetic field $B_{sw} =$ 1, 2, 3 and 4 G. The colorbars are the same for all maps, their scales are adjusted to highlight the magnetic field structure and do not reflect maximum and minimum values of $B$.
 }
   \end{minipage}
 \end{figure*}

This dependence is illustrated in Fig. \ref{fig:rMHD-PWN-vs-B} where the model magnetic field maps are shown for the meridional and equatorial cross-sections of the three-dimensional PWN model -- i.e., for the cross-sections by the plane formed by the pulsar spin axis and the local direction of $\textbf{B}_{sw}$ field (shown in white arrows) and the plane given by the pulsar's rotational equator.  These maps show that the magnetic field inside most of the volume of the PWN bubble inflated in the SW's Gauss-range magnetic field has sub-Gauss amplitudes. A strong Gauss-amplitude magnetic field  can be found in the equatorial outflows of the PWN -- fast magnetized relativistic outflows with velocities above $0.5c$ that carry plenty of even faster large-scale magnetic inhomogeneities. These inhomogeneities are very important for particle energizing by scatterings in the Fermi acceleration process in the colliding wind flows. Recall that in this process particles many times cross the contact discontinuity between two colliding flows, being upscattered by moving inhomogeneities in both of them, until their energy becomes too large to be confined in the colliding flows region.

\begin{figure}
\includegraphics[width=0.9\columnwidth]{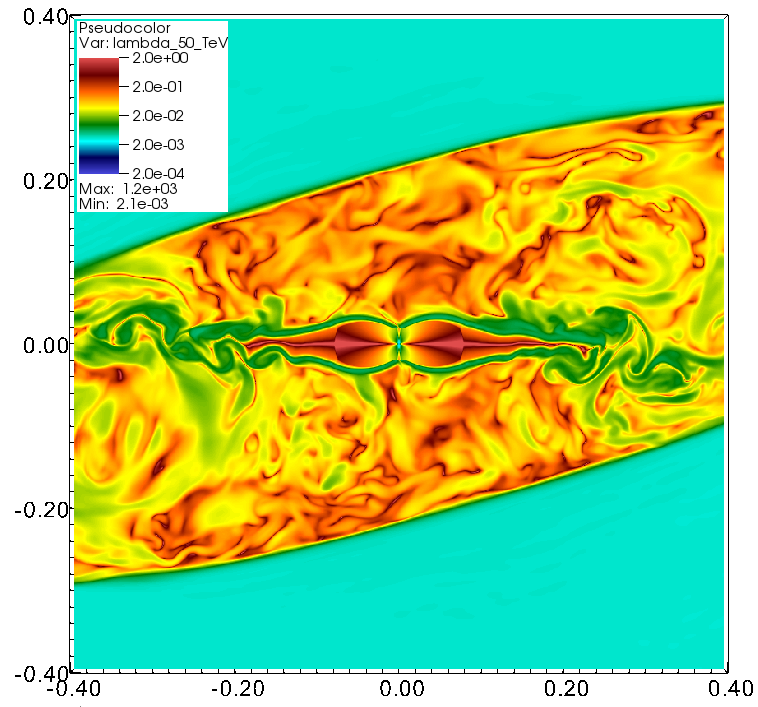} 
   \caption{Estimates of particle's mean free path (mfp) in the simulated structure of the bubble of a pulsar wind nebula blown in a dense magnetized stellar wind in a binary system. Shown is the map of mfp calculated in the Bohm diffusion approximation $\lambda = R_g = E / e B$ for particles with energy $E = 50$ TeV. The map is calculated using the simulated map of the magnetic field amplitude, the mfp in each point is computed using the local simulated value of $B$. The map shows a part of the meridional cut of the three-dimensional PWN model with the pulsar spin axis and stellar wind magnetic field lying in this plane (the former is directed upwards). Coordinates along the axes and the color-coded values of $\lambda$ are in AU.}
   \label{fig:leptons_mfp}
\end{figure}

The PWN bubble blown in strongly magnetized SW is bounded by a shell (or cocoon) of the perturbed SW matter. The enhanced SW magnetic field in this shell greatly contributes to the particles' acceleration, facilitating their confinement in the colliding winds zone. Meanwhile, particles with TeV-PeV energies have large mean free paths in the much lower magnetic field inside the PWN bubble, propagating  relatively freely in there (see Fig. \ref{fig:leptons_mfp}). As a result, particles of energies up to $E \sim E_{max}^{cwf}$, that may fall into the PeV range for protons \citep{2032our} and into TeV range for leptons (due to the radiative energy losses), populate the elongated bubble, producing  isotropic angular particle distribution inside its volume.

As the ultra-relativistic particles emit gamma-ray photons just along their velocity direction,  the elongated bubble populated with TeV particles would emit most of the TeV gamma-rays (regardless of being produced in the inverse Compton or p-p processes) along its large axis -- i.e., along the local large-scale magnetic field in the stellar wind $\textbf{B}_{sw}$. This would happen just because of a much larger thickness of the source in the direction along $\textbf{B}_{sw}$, rather than across it. Thus, if the SW's Gauss-range field is indeed characteristic for the GRLBs, a strong anisotropy of their VHE emission induced by this field could greatly impede their discovery, because most of their TeV luminosity would be detected only by the observer whose line of sight coincides with the $\textbf{B}_{sw}$. 

The anisotropy of the PWN bubble in the strongly magnetized SW is another possible reason why most of the Galactic binaries with pulsars may be missed in gamma-rays. This mechanism may be especially important for pulsar+Be-star binaries with intermediate and long orbital periods. For relatively large orbits, even small orbital eccentricity/inclination of the Be-disk discussed above can restrict the pulsar's residence inside the strongly magnetized inner regions of the wind -- and particularly inside the disk -- to a small fraction of the orbital period. This restriction limits the available range of directions where these objects may emit the VHE radiation to those swept by the major axis of a PWN bubble during the pulsar's passage through this fraction of the period.
For the pulsar+OB star binaries with shorter orbital periods and more uniform conditions along the orbit such effect may be less important.

\section{Leptons in a strong large-scale field. Anisotropy in energy losses}\label{sec:anisotropy-losses}

The inverse Compton radiation of accelerated leptons is thought to be responsible for the TeV emission of many GRLBs. As we already discussed in Sec. \ref{sec:strong-B-and-radiative-losses}, TeV leptons may experience severe radiative losses in these objects because of strong Gauss-range magnetic fields and intense radiation fields produced by the luminous massive companion. The analysis of the structure of magnetized flows in these objects, obtained in accurate modeling discussed in Sec. \ref{sec:anisotropic-PWNe}, shows that the influence of the radiative losses on the high energy emission of GRLBs is rather non-trivial.

\begin{figure}
    \begin{minipage}{1.0\columnwidth}
        \includegraphics[width=\textwidth]{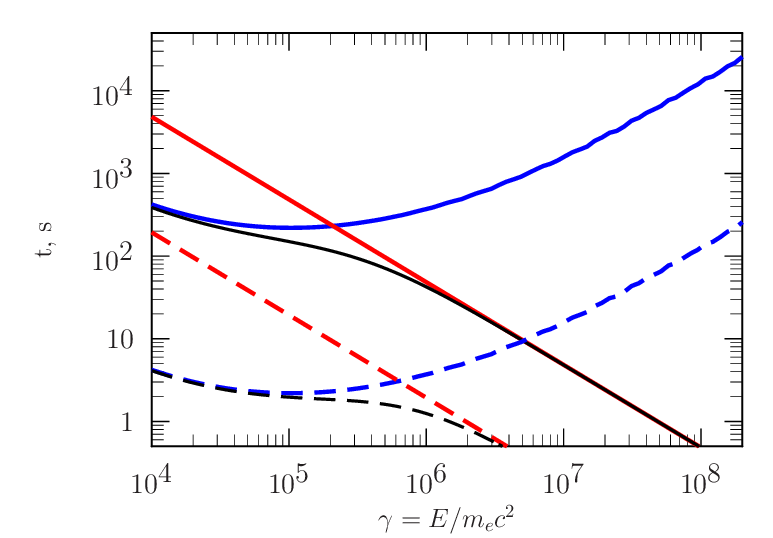} 
   \end{minipage}
   \caption{Typical times of radiative energy losses for leptons accelerated in the collision winds zone in a gamma-ray binary and propagating in the stellar wind. Shown is the dependence of loss times on the particle energy for synchrotron (red curves), inverse Compton (blue curves) and synchrotron + inverse Compton radiation (black curves). Note that estimates are calculated for averaged fields -- magnetic field and field of stellar radiation photons -- and do not take into account the direction of particle velocity with respect to their orientation. The parameters of calculation for solid curves: the magnetic field amplitude $B = 4 \:\mbox{G}$; the radiation field includes the cosmic microwave background and ultraviolet photons from a massive star. The radiation energy and density are calculated assuming a Planckian spectrum with $T = 3 \times 10^4$ K, the stellar radius 10 $R_{\odot}$, and the distance to the star $R =$ 1 AU. Dashed curves: the same, but for $B = 20$ G, $R = 0.1$ AU.
   }
\label{fig:t_cooling_leptons}
\end{figure}

The maximum energy $E_{max}$ that may be reached by leptons in a GRLB with a strong Gauss-range magnetic field was estimated in Sec. \ref{sec:strong-B-and-radiative-losses} by matching the timescales of particle acceleration and radiation losses. The estimate gives $E_{max} = 1$ TeV, assuming the magnetic field $B = B_{sw} = 4$ G in the particle acceleration zone. As discussed in Sec. \ref{sec:anisotropic-PWNe} and shown in Fig. \ref{fig:rMHD-PWN-vs-B}, in fact, the magnetic field inside the PWN bubble falls into a sub-Gauss range which may significantly boost the available $E_{max}$ value because of weaker synchrotron losses. However, in the following estimates we consider leptons with $E = 1 \:\mbox{TeV}$ and then briefly discuss the results for higher energies reached in acceleration.

\begin{figure*}
\begin{minipage}{1\textwidth}
   \begin{minipage}{.57\textwidth}
        \includegraphics[width=\textwidth]{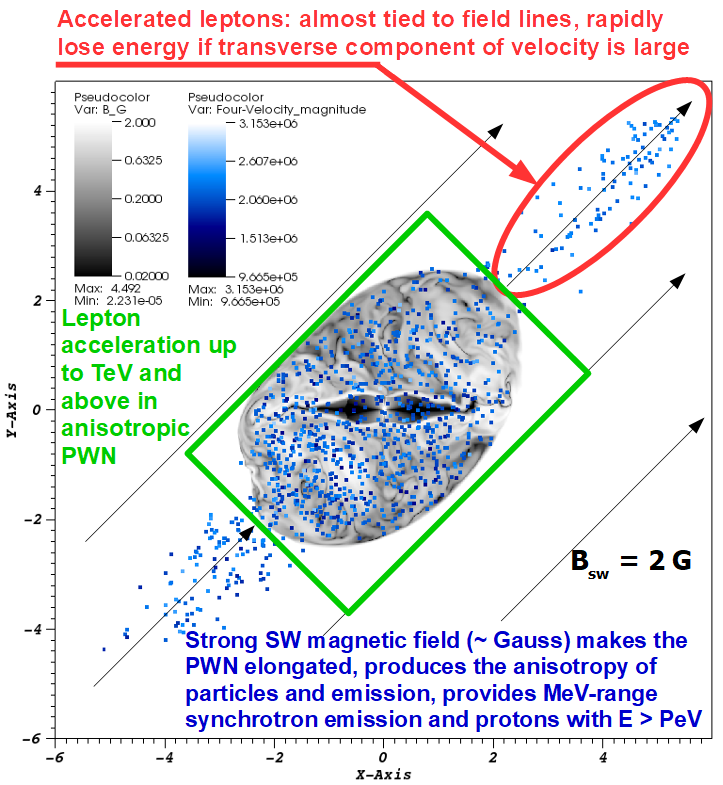}
   \end{minipage}
   \linebreak
   \vfill
   \begin{minipage}{0.8\textwidth}
     \includegraphics[width=\textwidth]{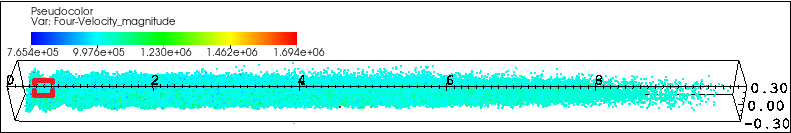}
   \end{minipage}
   \linebreak
   \vfill
   \begin{minipage}{.75\textwidth}
     \includegraphics[width=\textwidth]{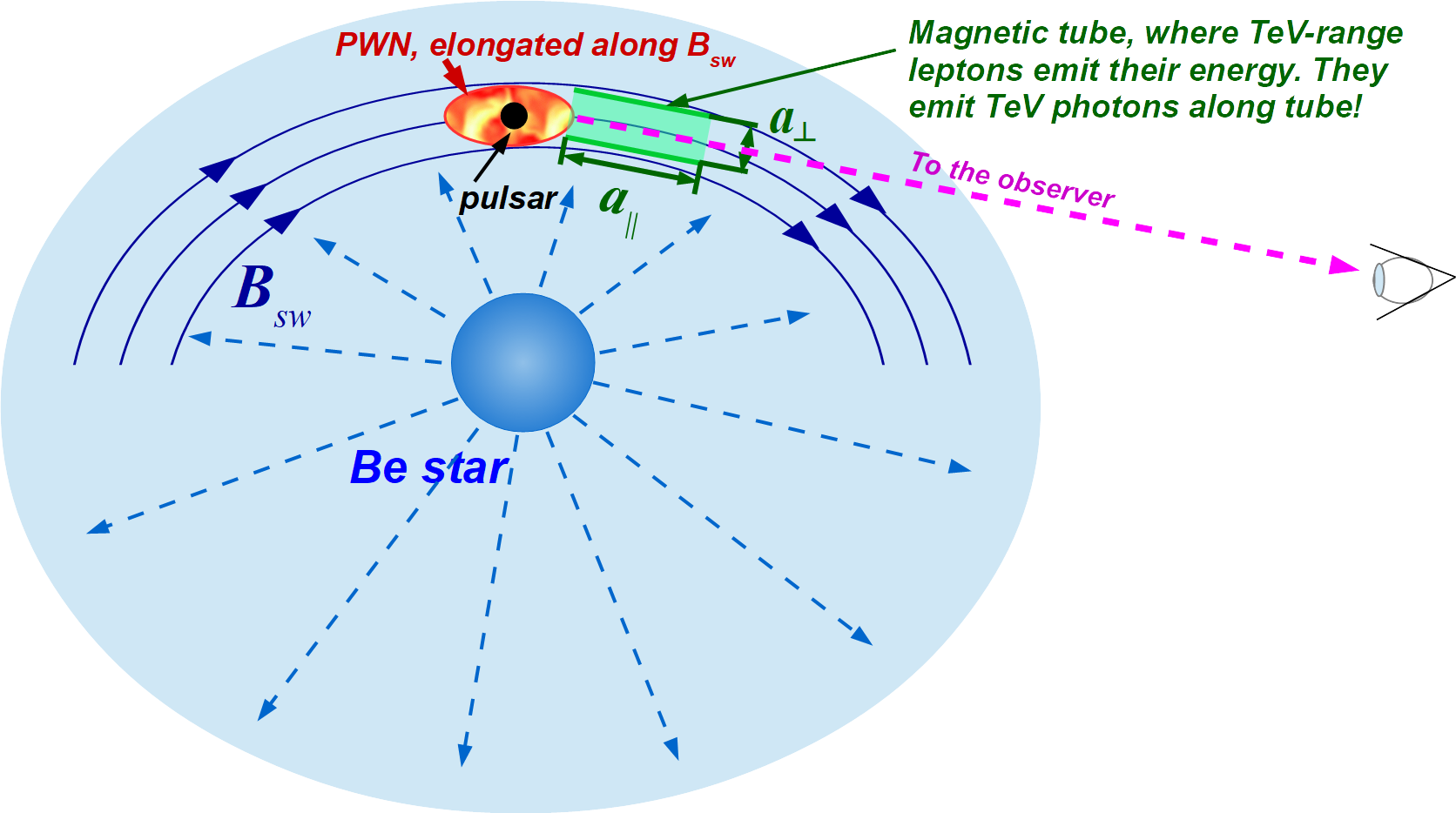}
   \end{minipage}
   \caption{Top: sketch of acceleration and escape from the accelerator of leptons accelerated in the magnetized structure of a PWN bubble inflated in a strongly magnetized stellar wind in a massive binary system. The results of rMHD-PIC simulation of particle transport are superimposed on the gray-color map of the simulated magnetic structure. Middle: the result of rMHD-PIC simulation of propagation of PeV protons along a strong large-scale field of Gauss range. Shown is the model spatial distribution of protons injected in a sub-AU-scale region (depicted in red) and propagating along a magnetic tube with strong large-scale magnetic field directed along the tube and multi-scale magnetic inhomogeneities. Bottom: sketch of production of a strongly anisotropic (sub)TeV radiation by the TeV-range leptons escaping from the PWN bubble in a PSR+Be system with a strong Gauss-range large-scale magnetic field in the equatorial disk.
\label{fig:sketch-leptons}}
 \end{minipage}
\end{figure*}

Consider leptons being accelerated in the colliding wind zone up to $E = 1 \:\mbox{TeV}$ which escaped the accelerator and propagate in a magnetized wind of the massive companion. This is schematically shown in the top panel of Fig. \ref{fig:sketch-leptons}, where the results of rMHD-PIC\footnote{In the framework of this approach particles are propagated using the conventional particle-in-cell (PIC) techniques in the structure of magnetized flows, simulated using the relativistic MHD model \cite{Mignone18}. This simulation, which is used just to illustrate the patterns of particle propagation in the collision winds zone of the GRLB with strong SW magnetic field, is analogous to ones shown in \cite{Bykov+24a}, though here the picture shows the meridional cut of a 3D rMHD simulation.} simulations of particle propagation in GRLB are illustrated. In the strong SW magnetic field, these leptons would be almost tied to the field lines and move along (or against) the field direction. Indeed, their gyroradius $R_g = E / e B_{sw} = 3 \times 10^9 \:\mbox{cm}\left(E / 1 \:\mbox{TeV}\right) \left(B_{sw} / 1 \:\mbox{G}\right)^{-1}$ is three orders of magnitude less than the typical spatial scales of the binaries. 
In the middle panel of this figure we show the results of kinetic modeling of the propagation of much more energetic -- PeV particles -- along such a field, taking into account the multiple-scale turbulent magnetic field of smaller scales capable of scattering the particles. This modeling shows that even PeV particles injected from the sub-AU accelerator propagate mostly along the field lines of a large-scale magnetic field with amplitude four G and produce a narrow tube-like ``magnetic jet'' (or ``magnetic tube''). A similar magnetic jet for TeV-range particles would be much better collimated.

In the strong magnetic field $B_{sw} = 4 \:\mbox{G}$ particles with a large velocity component transverse to $\textbf{B}_{sw}$ will rapidly lose their energy: Eq. (\ref{eq:synchro_time}) gives $t_{syn} \approx 17$ s for $\chi = \pi /2$. These particles will emit their energy in the keV-MeV range. Meanwhile, particles propagating almost along the $\textbf{B}_{sw}$ with $\sin^2\chi \ll 1$ lose energy via synchrotron radiation much slower, and thus may convert their energy in (sub)TeV photons detectable by {\sl H.E.S.S.}

The length $a_{\parallel}$ at which particles with energy $E$ propagating almost along $\textbf{B}_{sw}$ will lose energy due to the inverse Compton radiation is given by $a_{\parallel} = ct_C\left(E\right)$, where the typical time of energy losses due to the inverse Compton radiation $t_C$ is given by blue curves in Fig. \ref{fig:t_cooling_leptons}. For $E \gsim$ 1 TeV it gives $a_{\parallel} \sim 1$ AU, which is rather small compared with the inferred sizes of Be-star disks that may reach $400 R_{\ast} \sim 20$ AU \citep{Klement+17}. Thus, in the medium- and long-period GRLBs with Be-stars, where the typical orbital distance between the pulsar and the massive star is $\gsim 1$ AU, particles of TeV energies escaping from the AU-scale accelerator 
would populate an almost straight magnetic jet of AU-scale length. This magnetic jet would be a strongly anisotropic emitter of (sub)TeV gamma-rays, because these particles would emit photons almost along their velocity direction -- i.e. almost along the $\textbf{B}_{sw}$, as  shown in the bottom panel of Fig. \ref{fig:sketch-leptons}.

Let us estimate which fraction of leptons escaping the accelerator would contribute to the TeV emission of the magnetic jet. These are leptons whose pitch-angle $\chi$ is small enough to make the synchrotron losses slower than the inverse Compton ones. We are looking for the maximum angle $\chi_0$ allowing a particle to propagate along the magnetic field lines at a distance of $\sim$ 1 AU without significant synchrotron losses. Equation (\ref{eq:synchro_time})
gives for $B = 4 \:\mbox{G}$ and $E = 1$ TeV the desired $c t_{syn} \approx 1.5 \times 10^{13} \:\mbox{cm}$ for $\sin\chi_0 \approx 0.2$, or $\chi_0 \approx 11^{\circ}$.

Let us now estimate the luminosity of this anisotropic TeV emitter. Let us assume that the PWN converts 1\% $\dot{E}$ into TeV particles -- and $\eta \dot{E} = 0.005 \dot{E}$ is injected to our emitter. We also assume that the injected particles are distributed uniformly over $\mu_{p} = \cos\chi$ (ranging from 0 to 1) and the azimuthal angle $\phi$ (ranging from 0 to 2$\pi$). Then the power emitted in (sub)TeV photons $L = \left(\Delta\Omega / 4\pi\right)\cdot \eta \dot{E} \approx (\chi_0^2 / 4) \eta \dot{E}\approx 5 \times 10^{-5} \dot{E}$, where $\Delta \Omega = 2 \pi (1-\cos\chi_0)$ is the solid angle subtended by a cone with the half-angle $\chi_0$. Assuming that the photons propagate along the magnetic field and the width of the magnetic jet is $a_{\perp}$, the energy flux through the distant (from the pulsar) base of the magnetic tube is $\sim L / a_{\perp}^2$. Then the observer at distance $d$ will detect a $\left(a_{\parallel} /d\right)^2$ times lower flux. For $\dot{E} = 1.7 \times 10^{35} \ergs$ and $d = 1.3$ kpc (parameters of PSR J2032+4127 \citep{Ho+17}) one obtains $F \sim 5 \times 10^{-13} \left(a_{\parallel} / a_{\perp}\right)^2 \:\mbox{erg}\,\mbox{cm}^{-2}\,\mbox{s}^{-1}$. Assuming that $a_{\perp}$ is of the order of the PWN bubble transverse (to the SW magnetic field) size, which is according to Fig. \ref{fig:rMHD-PWN-vs-B} has a sub-AU scale, one may readily expect  $F > 10^{-12} \:\mbox{erg}\,\mbox{cm}^{-2}\,\mbox{s}^{-1}$ from such objects in (sub)TeV range.

\begin{table}[h!]
    \centering
    \begin{tabular}{lcccc}
    \hline       
    $E_{max}$, TeV & & $\chi_0$,$^{\circ}$ & & $F$, $10^{-12}$ erg\,cm$^{-2}$\,s$^{-1}$ \\
    \hline
    1 & & 11 & & 2.1 \\
    2 & & 7.9 & & 1.1 \\
    4 & & 5.6 & & 0.53 \\
    10 & & 3.5 & & 0.21
    \end{tabular}
    \caption{Estimates of (sub)TeV emission flux and half-opening angle of the cone defining the TeV radiation pattern of a magnetic jet producing the emission. The jet is filled by TeV particles accelerated in a PWN bubble and propagating along a strong magnetic field of the stellar wind. The flux is estimated assuming the distance $d = 1.3$ kpc, pulsar spin-down energy rate $\dot{E} = 1.7 \times 10^{35} \ergs$ and the fraction of spin-down power converted into the TeV particles $\eta = 10^{-2}$} 
    \label{table:lepton-anis-emitter}
\end{table}

In Table 2 we present the estimates of $\chi_0$ and $F$ for somewhat larger values of $E_{max}$, assuming $a_{\perp} = 0.5$ AU. Higher energies imply stronger synchrotron loss rates and, therefore, stronger anisotropy of the considered emitter. These results show that the emission produced by TeV-range leptons accelerated in the Be+PSR gamma-ray binaries may be rather bright, but detectable only in a narrow range of angles.

The absorption of high-energy $\gamma$-rays due to collisions with stellar radiation  \citep{Nikishov1962} may substantially reduce the $\gamma$-ray luminosity produced in a close vicinity of a young massive star \citep[see e.g.][]{Dubus08,prediction-nu-Neronov2009,takata17}. A strong anisotropy of high-energy radiating particles due to the magnetization of stellar wind discussed above should affect the optical depth for TeV $\gamma$-ray interaction with the stellar and disk photons, making the absorption effect strongly geometry dependent. In our case the direction of the local magnetic field is an important geometrical factor to be included in calculations of the optical depth for high-energy $\gamma$-rays.



\section{Discussion}

In this work, we considered the results of the population synthesis simulations of the expected number of Galactic binary systems with colliding winds from a young pulsar and a young massive star which are expected to be VHE gamma-ray sources. The analysis by Dubus et al. \cite{2017A&A...608A..59D} suggested that the number of gamma-ray binaries in our Galaxy that may be detected in HE and VHE surveys can be estimated as 101$^{+89}_{-59}$. 
\textbf{Motivated by the current detection of PeV photons by LHAASO, we mainly focused here on studying in some detail a possible number of the VHE gamma-ray binaries with young pulsars which can be sources of PeV cosmic rays. Two main questions we address here include (i) calculation of the expected number
of potential VHE particle accelerators among the binaries with young pulsar and (ii) discussion of their observability by Cherenkov observatories as gamma-ray binaries.}

The simulated binary statistics predicts about two dozen of the sources with characteristic spin-down power of the pulsar  $\dot{E}_{35} \gsim 1$ which exceeds the number of detected GRLBs. \textbf{We caution, however, that population synthesis calculations are model-dependent (CE treatment, binary parameter distributions and fraction, star formation rate normalization, supernova kicks, etc.) and therefore give approximate numbers of corresponding binary Galactic populations to within at least a factor of two.}
Two observed pulsars PSR B1259–63 and PSR J2032+4127 with long orbital periods demonstrated bright VHE emission mostly during the periastron passage which indicates that the VHE radiation is likely originated due to pulsar wind interaction with an anisotropic outflow from the massive star. Our modeling showed that a strong Gauss-range magnetic field in the stellar wind as inferred to explain a number of high-energy spectral features in these objects is a key parameter to accelerate VHE particles in pulsar-driven GRLBs \citep{Bykov+24a}.

The VHE particle acceleration according to  Eq. (\ref{eq:E_max_aniso_outflow}) is possible, in principle, even for binary systems with $\dot{E}_{35} \gsim 0.01$ assuming the pulsar wind opening angle 
$\Omega \lsim 0.1$. The predicted number of Be+PSR binaries with $\dot{E}_{35} \gsim 0.01$ is about 100 (see Fig. \ref{fig:Be-OB-distr}).
The effect of the HE radiation anisotropy (beaming of the released energy in a cone) was proposed to explain the bright GeV flares detected by the Fermi LAT telescope around the periastron passage by PSR B1259–63 in 2017 \cite{2018ApJ...863...27J,ChernyakovaGeV20} and 2021  \cite{Chernyakova_univ21}. The detected GeV  outbursts
require the isotropic gamma-ray luminosity of the source to exceed  the estimated  spin-down luminosity of PSR B1259-63 (which is about $8 \times 10^{35} \ergs$) by a factor of $\sim$ 30 for the 2017 periastron passage  and  $\sim$ 6 for 2021. Even for a high enough conversion efficiency of the spin-down luminosity to GeV radiation, the degree of anisotropy of the emission must be rather strong. The GeV radiating cone can be naturally produced if the pulsar wind is interacting with the highly magnetized stellar wind as clearly seen in Fig. \ref{fig:rMHD-PWN-vs-B}. The cone opening angle and its direction will evolve along the pulsar orbit following the local magnetic field variations. Spectropolarimetric VLT measurements of the mean longitudinal magnetic fields in the stellar companion of VHE emitting binary LS 5039 \cite{2019IAUS..346...40H,2021mfob.book.....H} reported a detection of the field with the estimated magnitude $794 \pm 277$ G
at a significance level of 2.5 $\sigma$. If confirmed, this can provide a Gauss-range magnetic field in the interaction region of the compact object and stellar wind. The presence of a Gauss-range magnetic field in the LS 5039 periastron region was suggested earlier in the HE spectral modeling of LS 5039 \cite{Dubus08} and it was attributed to the possible high magnetization parameter of the pulsar wind in \cite{2015A&A...581A..27D} to explain the MeV detection by COMPTEL \cite{Collmar14}. Also, multi-zone modeling of LS 5039 presented in \cite{2015A&A...575A.112D} required moderate magnetic field magnitudes of $\lsim$ G in the radiation zone to fit the HE data.        

The Gauss-range magnetic fields give rise to a strong anisotropy of the VHE emission of GRLBs which alleviates the system power requirements to allow $\dot{E}_{35} \gsim 0.01$. The simulated number of such Be-pulsar binaries in Fig. \ref{fig:Be-OB-distr} is about 100.  However, the strong anisotropy may limit the number of detections of the objects (especially with long orbital periods) in gamma-rays to the observer's viewing angles coaligned with the magnetic field direction. 
Moreover, the VHE luminosities of such objects are likely to be smaller than that of PSR B1259–63 and PSR J2032+4127. Future high sensitivity VHE observations will be able to find or constrain the lower luminosity GRLBs. The population of a hundred of VHE binary sources with  $\dot{E}_{35} \gsim 0.01$ can provide the fluxes of PeV energy cosmic rays just comparable to that of a very few GRLBs with $\dot{E}_{35} \gsim 1$. However, the observed low anisotropy of the galactic cosmic rays which is well below 1\% at energies $\sim$ PeV is not easy to explain by the presence of a very few PeV cosmic ray sources. Therefore, a large population with $\sim$ 100 weaker sources is likely needed indeed. While the total power of the VHE binary sources with young pulsars under discussion is just above $10^{36} \ergs$, which is much less than that of the microquasar type binaries like SS 433 (see e.g. \citep{Churazov+24}) powered by supercritical accretion onto a black hole, still their duty cycles may be much longer than the microquasar activity phase, thus possibly providing a comparable integral contribution.   
Recently, the LHAASO Observatory measured the cosmic ray proton fluxes at PeV energies \cite{2025arXiv250514447T} and also detected the diffuse gamma-ray emission from 10 TeV to 1 PeV from the Galactic Plane \textbf{where we expect to find young pulsars} \cite{2023PhRvL.131o1001C}. The data being confronted with the current cosmic ray models using the global cosmic-ray  propagation codes in \cite{2025arXiv250606593E} show some mismatch 
indicating the need of further study of sources of PeV-energy cosmic rays.    

\section*{CRediT authorship contribution statement}
\textbf{A. M. Bykov}: Conceptualization, Methodology, Software–simulation of GRLBs structure, Formal analysis, Investigation, Writing – original draft,  Writing – review \& editing, Project administration. \textbf{A. G. Kuranov}: Software–simulation of GRLBs population synthesis, Formal analysis, Investigation, Writing – original draft, Writing – review \& editing. \textbf{A. E. Petrov}: Software – calculation of VHE particles propagation in GRLBs, Formal analysis, Investigation, Writing – original draft, Writing – review \& editing. \textbf{K. A. Postnov}: Conceptualization, Methodology, Software–simulation of GRLBs population synthesis, Formal analysis, Investigation, Writing – original draft, Writing– review \& editing.

\section*{Declaration of competing interest}
The authors declare that they have no known competing financial interests or personal relationships that could have appeared to influence the work reported in this paper.

\section*{Data availability}
The data will be made available on request.

\section*{Acknowledgements}
We thank the anonymous referee for careful reading of our manuscript and asking a number of important and  interesting questions  
that allowed us to improve the presentation of our results.
Simulations of massive binary statistics by AGK was supported by RSF grant No. 25-22-00295 (population synthesis of Galactic massive binaries with pulsars). Modeling of the pulsar wind nebulae structure in magnetized stellar winds by AMB  with the  Lomonosov-2 computer facilities was supported by RSF grant No. 25-72-20007 (Multiscale nonlinear models of astrophysical sources of high energy radiation). Modeling of very high energy particle propagation in gamma-ray binaries by AEP was supported by the Foundation for the Advancement of Theoretical Physics and Mathematics ``BASIS'' (grant No. 24-1-3-28-1). It was performed using the resources of the supercomputer center of St. Petersburg Polytechnic University, http://scc.spbstu.ru. 
%












\end{document}